\documentclass[12pt,preprint]{aastex}
\usepackage{natbib,times}
\usepackage{lscape}
\usepackage{xcolor}

\newcommand{\nuc}[2]{$\mrm{^{#2}#1}$}

\newcommand{\mrm}[1]{\mathrm{#1}}

\def\beq{\begin{equation}}
\def\enq{\end{equation}}
\def\ms{$M_\odot$}
\def\Al{$^{26\!}$Al\ }
\def\Fe{$^{60\!}$Fe\ }
\shorttitle{}
\shortauthors{}
\begin{document}
\title{Gamma-ray Emission of \Fe and \Al Radioactivities in our Galaxy}

\author{W. Wang$^{1,2}$, T. Siegert$^{3,4}$, Z. G. Dai$^{1,5}$, R. Diehl$^{4,6}$, J. Greiner$^{4}$, A.~Heger$^{7,8}$, M. Krause$^9$, M. Lang$^4$, M. M. M. Pleintinger$^4$, X.L. Zhang$^4$ }
\affil{$^1$ School of Physics and Technology, Wuhan University, Wuhan 430072, China; wangwei2017@whu.edu.cn \\
       $^2$ WHU-NAOC Joint Center for Astronomy, Wuhan University, Wuhan 430072, China \\
       $^3$ Center for Astrophysics and Space Sciences, UC San Diego,  92093-0424, La Jolla (CA), U.S.A.; tsiegert@ucsd.edu \\
       $^4$ Max-Planck-Institut f\"{u}r extraterrestrische Physik,  85741 Garching, Germany; rod@mpe.mpg.de  \\
       $^5$ School of Astronomy and Space Science, Nanjing University, Nanjing 210093, China; dzg@nju.edu.cn \\
       $^6$ Munich Cluster of Excellence `Universe', 85748 Garching, Germany  \\
       $^7$ School of Physics and Astronomy, Monash University, VIC 3800, Australia \\
       $^8$ Tsung-Dao Lee Institute, Shanghai 200240, China \\
       $^9$ Centre for Astrophysics,  University of Hertfordshire, Hatfield, AL10 9AB, United Kingdom\\
       }

\begin{abstract}
The isotopes $^{60}$Fe and $^{26}$Al originate from massive stars and their supernovae, reflecting ongoing nucleosynthesis in the Galaxy. We studied the gamma-ray emission from these isotopes at characteristic energies 1173, 1332, and 1809\,keV with over 15 years of SPI data, finding a line flux in \Fe combined lines of $(0.31\pm0.06) \times 10^{-3}\,\mrm{ph\,cm^{-2}\,s^{-1}}$ and the \Al line flux of $(16.8\pm 0.7) \times 10^{-4}\,\mrm{ph\,cm^{-2}\,s^{-1}}$ above the background and continuum emission for the whole sky. Based on the exponential-disk grid maps, we characterise the emission extent of \Al to find scale parameters $R_0 =7.0^{+1.5}_{-1.0}$ kpc and $z_0=0.8^{+0.3}_{-0.2}$ kpc, however the \Fe lines are too weak to spatially constrain the emission. Based on a point source model test across the Galactic plane, the \Fe emission would not be consistent with a single strong point source in the Galactic center or somewhere else, providing a hint for a diffuse nature. We carried out comparisons of emission morphology maps using different candidate-source tracers for both \Al and \Fe emissions, and suggests that the \Fe emission is more likely to be concentrated towards the Galactic plane. We determine the \Fe/\Al $\gamma$-ray flux ratio at $(18.4\pm4.2)\,\%$ , when using a parameterized spatial morphology model. Across the range of plausible morphologies, it appears possible that \Al and \Fe are distributed differently in the Galaxy. Using the best fitting maps for each of the elements, we constrain flux ratios in the range 0.2--0.4. We discuss its implications for massive star models and their nucleosynthesis.

\end{abstract}


\section{Introduction}
The radioactive isotope $^{60}$Fe is produced in suitable astrophysical environments through successive neutron captures on pre-existing Fe isotopes such as (stable)
$^{54,56,57,58}$Fe in a neutron-rich environment. Candidate regions for \Fe production are the He and C burning shells inside massive stars, where neutrons are likely to be released from the $^{22}$Ne($\alpha$,n) reaction. \Fe production may occur any time during late evolution of massive stars towards core collapse supernovae (Woosley \& Weaver 1995; Limongi \& Chieffi 2003, 2006, 2013; Pignatari et al. 2016; Sukhbold et al. 2016; and references therein). There is also an explosive contribution to the \Fe yield by the supernova shock running through the carbon and helium shells (Rauscher et al. 2003). Electron-capture supernovae may be a most-significant producer of $^{60}$Fe in the Galaxy (Wanajo et al. 2013, 2018; Jones et al. 2016, 2019a). There are other possible astrophysical sources of $^{60}$Fe. From similar considerations, $^{60}$Fe can also be made and released in super-AGB stars (Lugaro et al. 2012). Furthermore, high-density type Ia supernova explosions that include a deflagration phase (Woosley 1997) can produce even larger amounts per event.

Due to its long lifetime (radioactive half-life $T_{1/2} \simeq$2.6~Myr, Rugel et al. 2009; Wallner et al. 2015; Ostdiek et al. 2017), $^{60}$Fe survives to be detected in $\gamma$-rays after being ejected into the interstellar medium: \Fe $\beta$-decays to $^{60}$Co, which decays within 5.3~yr to $^{60}$Ni into an excited state that cascades into its ground state by $\gamma$-ray emission at $1173$ keV and 1332 keV. $^{26}$Al has a similarly-long radioactive lifetime of $\sim 10^{6}$ years, and had been the first live radioactive isotope detected in characteristic $\gamma$-rays at 1809~keV (Mahoney et al. 1978), thus proving currently ongoing nucleosynthesis in our Galaxy. Mapping the diffuse $^{26}$Al $\gamma$-ray emission suggested that it follows the overall Galactic massive star population (Diehl et al. 1995; Prantzos and Diehl 1996). If $^{60}$Fe and $^{26}$Al have similar astrophysical origins, with their similar radioactive life time, the ratio of $^{60}$Fe to $^{26}$Al $\gamma$-ray emissions in the Galaxy would be independent of the true specific distances and locations of the sources, and thus important for testing stellar evolution models with their nucleosynthesis and core-collapse supernova endings (Woosley and Heger 2007; Diehl 2013).

The steady-state mass of these radioactive isotopes maintained in the Galaxy through such production counterbalanced by radioactive decay thus converts into a ratio for the $\gamma$-ray flux in each of the two lines through
\begin{equation}
\frac{ I(^{60}Fe)}{ I(^{26}Al)} \sim 0.43 \cdot \frac{\dot{M}(^{60}Fe)}{\dot{M}(^{26}Al}
\label{eq:fealfluxratio}
\end{equation}
Timmes et al. (1995) carried the massive-star yields into an estimate of chemical evolution for $^{60}$Fe and $^{26}$Al in the Galaxy, predicting a $\gamma$-ray flux ratio of 0.16. Various revisions of models presented different ratio values ($\sim 0.5 - 1$, see Limongi \& Chieffi 2003, 2006; Rauscher et al. 2002; Prantzos 2004; Woosley and Heger 2007).  Different massive star regions, such as Scorpius-Centaurus or Cygnus, may show a different \Fe/\Al ratio as the age of such associations dictates the expected fluxes (Voss et al. 2009). The average over the whole Milky Way, on the other hand, is determined by the number of massive stars and their explosions over the characteristic lifetimes of \Al and \Fe, respectively, providing an independent measure of the core-collapse supernova rate in the Milky Way, and up to which masses stars actually explode.

The yields of these two isotopes depend sensitively on not only the stellar evolution details such as shell burnings and convection, but also the nuclear reaction rates. Tur et al. (2010) found that the production of $^{60}$Fe and $^{26}$Al is sensitive to the 3$\alpha$ reaction rates during He burning, i.e. the variation of the reaction rate by a factor of two will make a factor of nearly ten change in the $^{60}$Fe/$^{26}$Al ratio. \Fe may be destroyed within its source by further neutron captures \Fe($n,\gamma$). Since its closest parent, $^{59}$Fe is unstable, the
$^{59}$Fe($n,\gamma$) production process competes with the $^{59}$Fe $\beta$ decay to produce an appreciable amount of \Fe. This reaction pair dominates the nuclear-reaction uncertainties in \Fe production, with $^{59}$Fe($n,\gamma$) being difficult to measure in nuclear laboratories due to its long lifetime and E1 and M1 reaction channels (Jones et al. 2019b).
Using effective He-burning reaction rates can account for correlated behaviour of nuclear reactions and mitigate the overall nuclear uncertainties in these shell burning environments (Austin et al, 2017). Astronomical observations of the \Fe/\Al ratio will help to constrain the nuclear-reaction aspects of massive stars, given the experimental difficulties to measure all reaction channels involved at the astrophysically-relevant energies.

Such comparison and interpretation, however, relies on the assumption that \Al and \Fe originate from the same sources (see, e.g., Timmes et al. 1995; Limongi \& Chieffi 2006, 2013), and have the similar diffuse emission distributions, originating from nucleosynthesis in massive stars throughout the Galaxy. Therefore, the observational verification of the diffuse nature of \Fe emission is key to above interpretation of measurements of a \Fe/\Al ratio in terms of massive-star models.

Several detections of \Fe enriched material in various terrestrial as well as lunar samples (Knie et al. 2004; Wallner et al. 2015; Fimiani et al. 2016; Neuh\"auser et al. 2019) confirm the evidence for a very nearby source for \Fe within several Myr. A signal from interstellar \Fe was first reported from the NaI spectrometer aboard the RHESSI spacecraft (2.6$\sigma$) which was aimed at solar science (Smith 2004). They also presented a first upper limit of $\sim 0.4$ (1$\sigma$) for the flux ratio of $^{60}$Fe/$^{26}$Al $\gamma$-ray emissions (Smith 2004).
The first solid detection of Galactic \Fe emission was obtained from INTEGRAL/SPI measurements, detecting \Fe $\gamma$-rays with a significance of $4.9\sigma$ after combining the signal from both lines at 1173 and 1332 keV (Wang et al. 2007), constraining the flux ratio of $^{60}$Fe/$^{26}$Al $\gamma$-ray emissions to the range of $0.09- 0.21$ (Wang et al. 2007). Subsequent analysis of ten-year INTEGRAL/SPI data with a different analysis method similarly suggested a ratio in the range $\sim 0.08-0.22$ (Bouchet et al. 2015).

In this paper, we use more than 15 years of SPI observations through the entire Galaxy, and carry out a broad-band spectral analysis in the $\gamma$-ray range 800 keV -- 2000 keV. Rather than striving for high spectral resolution and line shape details, this wide-band $\gamma$-ray study aims at study of both \Fe and \Al emission signals at 1173, 1332 and 1809\,keV simultaneously, i.e., using identical data and analysis methods, including data selection and background treatments. This paper is structured as follows. In \S\,\ref{sec:dataobs}, we will describe the SPI observations and the data analysis steps. Our emission models to describe the 800--2000\,keV band are introduced in \S\,\ref{sec:skymodels}. We present our morphological as well as spectral findings in \S\,\ref{sec:results}. Implications and conclusion are shown in Section \S\,\ref{sec:discussion}.

\section{Observations and data analysis}\label{sec:dataobs}

\subsection{SPI observational data}

The {\em INTEGRAL\/} mission (Winkler et al. 2003) began with its rocket launch on October 17, 2002.
The spectrometer SPI  (Vedrenne et al. 2003) is one of INTEGRAL's two main telescopes.
It consists of 19 Ge detectors, which are encompassed into a BGO detector system used in anti-coincidence for background suppression.
SPI has a tungsten coded mask in its aperture, which allows imaging with a $\sim 3^{\circ}$ resolution
within a $16^{\circ} \times 16^{\circ}$ fully-coded field of view. The Ge detectors record $\gamma$-ray events from energies between 20 keV and 8 MeV.
The performance of detectors, and the behaviour and variations of instrumental backgrounds, have been studied over the mission, and confirmed that scientific performance is maintained throughout the mission years (Diehl et al. 2018). The {\em INTEGRAL} satellite with its co-aligned instruments is pointed at predesignated target regions, with a fixed
orientation for intervals of typically $\sim 2000$ s (referred to
as {\em pointings\/}).

The basic measurement of SPI consists of event messages per photon
triggering the Ge detector camera. We distinguish events which
trigger a single Ge detector element only (hereafter {\it single
event}, SE), and events which trigger two Ge detector elements
nearly simultaneously (hereafter {\it multiple event}, ME). The fast Pulse Shape Discriminator electronic unit (PSD), digitizes the shape of the current pulse, and allows suppression of background events, e.g. from localized $\beta$-decays within the Ge detectors (Roques et al. 2003), or from electronic noise. In this work, we use event data which hit only one detector (i.e., SE event data) and which carry the PSD flag for acceptable pulse shape (event type `PE').

We applied a selection filter to reject corrupted, invalid, or background-contaminated
data. We apply selection windows to `science housekeeping'
parameters such as the count rates in several background-monitoring detector rates,
proper instrument status codes, and orbit phase. In particular, we use the SPI plastic scintillator anti-coincidence
counter (PSAC) mounted beneath the coded mask, and the rate of saturating events in SPI's Ge
detectors (i.e., events depositing $>$ 8 MeV in a single Ge detector; hereafter referred to as GeDSat rates) as background tracers. This selection leads to exclusions of strong solar-flare periods and other times of clearly increased / abnormal backgrounds. Additionally, regular
background increases during and after passages through the Earth's
radiation belts are eliminated by a 0.1--0.9 window on orbital phase.
As a final step after these primary selections, we perform a quality-of-fit selection using our best background model only (see \S\,\ref{sec:dataanalysis}), and exclude all further pointings which show deviations beyond $10\ \sigma$ above  a fit of this background plus the expected signal contribution per pointing,
thus eliminating pointings with abnormal background (note that SPI data are dominated by instrumental background counts, so that source counts from diffuse emission such as \Fe alone cannot deteriorate the fit to a pointing dataset significantly). These outliers are mainly due to missing `science housekeeping' parameters which are interpolated later, or due to burst like events in the field of view, for example gamma-ray bursts, flares from X-ray binaries, or solar activity.

\begin{figure}
\centering
\includegraphics[angle=0,width=10cm]{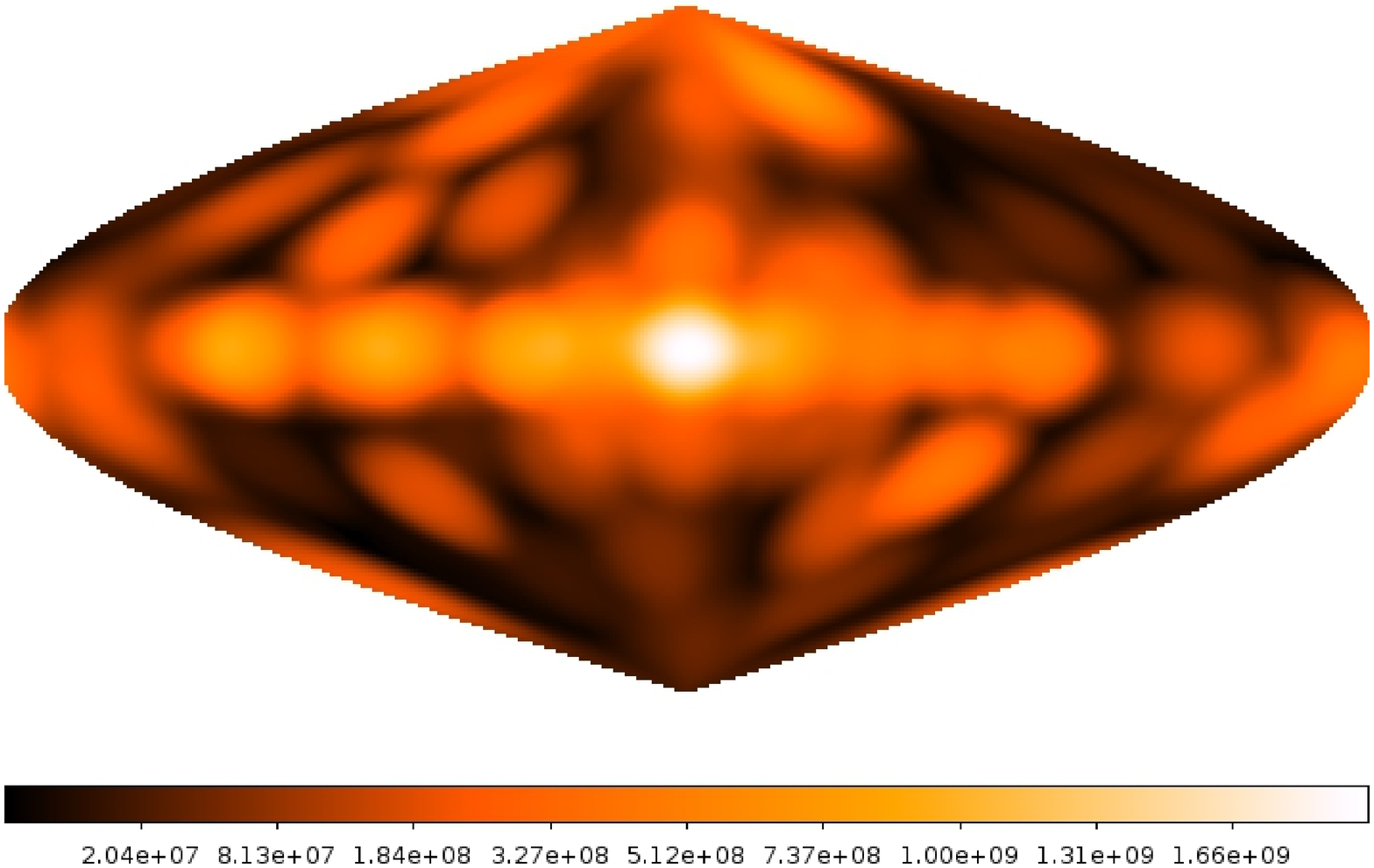}
\caption{Exposure sky map of fully-coded FOV in Galactic coordinates (the number
at the color bar in units of sec) for the data selected from
15-year SPI observations for our \Fe and \Al study (INTEGRAL orbits 43 --
1950).}
\label{fig:exposuremap}
\end{figure}

The resulting dataset for our \Fe and \Al study encompasses 99,864 instrument pointings across the entire sky, equivalent
to a total (deadtime-corrected) exposure time of 213\,Ms. This includes data from INTEGRAL orbits 43 to 1950, or February 2003 to May 2018.
The sensitivity (effective exposure time, effective area) of SPI observations is further reduced by the successive failure of four of the 19 detectors (December 2003, July 2004, 19 Feb 2009 and 27 May 2010). In Fig.\,\ref{fig:exposuremap} we show the resulting exposure map from our cleaned data set.

We bin detector events from the range between 800\,keV and 2000\,keV into seven energy bins: three of these address $\gamma$-ray line bands for \Fe and \Al (1169 -- 1176\,keV; 1329 -- 1336\,keV; 1805 -- 1813\,keV), and four continuum bands in between (800 -- 1169\,keV; 1176 -- 1329\,keV; 1336 -- 1805\,keV; 1813 -- 2000\,keV) are added to robustly determine the line flux above the diffuse $\gamma$-ray continuum. In the Appendix, we present an investigation of the impact of event
selections using the PSD (see Fig.\,\ref{fig:sepsdcomparison}). PSD selections succeed to suppress an
apparent electronics noise that is visible in raw detector spectra in
the energy range 1300 --1700 keV. We therefore use such PSD selections,
although the impact on the resulting spectra from celestial sources is
not clear except for this continuum band (see Appendix).

\subsection{Data analysis}\label{sec:dataanalysis}

The raw data of SPI are dominated by the intense background (BG) radiation
characteristic of \hyphenation{space} platforms undergoing cosmic-ray (CR) bombardment.
In SPI data analysis, we combine background models with a spatial model for sky
emission to fit our data, allowing for adjustments of
fit parameters for background and sky intensities.
In general, the counts per energy bin, per detector, and per
pointing are fitted by the background model described in Siegert et al. (2019), with details outlined shortly below, and the assumed sky map of celestial emission (e.g., the \Al
distribution obtained by COMPTEL or exponential disk models, see \S\,\ref{sec:skymodels}) as convolved into
the domain of the SPI data space for the complete pointing sequence, by the instrument coded-mask response:

\beq D_{e,d,p}=\sum_{m,n}\sum_{j=1}^{k_1}
A_{e,d,p}^{j,m,n}\beta_s^j I_j^{m,n} + \sum_t
\sum_{i=1}^{k_2}\beta^i_{b,t} B^i_{e,d,p} + \delta_{e,d,p}, \label{eq:datafit}\enq
where {\it e,d,p} are indices for data space dimensions: energy,
detector, pointing; {\it m,n} indices for the sky dimensions
(galactic longitude, latitude); $A$ is the instrument response
matrix, $I$ is the intensity per pixel on the sky, $k_1$ is the number of independent sky intensity distribution
maps; $k_2$ is the number of background components, $\delta$ is
the count residue after the fitting. The coefficients $\beta_s$ for the sky map intensity are constant in time, while
$\beta_{b,t}$ is allowed time dependent, see \ref{sec:bg_char} below.
The sky brightness amplitudes $\beta_s$ comprise the resultant spectra of the signal
from the sky. For this model fitting analysis, we use a maximum-likelihood method, implementing Poisson statistics which applies to such detector count analysis.
Our software implementation is called {\em spimodfit} (Strong et al. 2005).
The fitted model components are analysed for further consistency checks on possible systematics in residuals.

We thus derive, per energy bin, best-fitted parameter
values with uncertainties, and their covariance matrices.
This provides flux estimates which are independent of spectral-shape expectations.

\begin{figure}
	\centering
	\includegraphics[width=9cm]{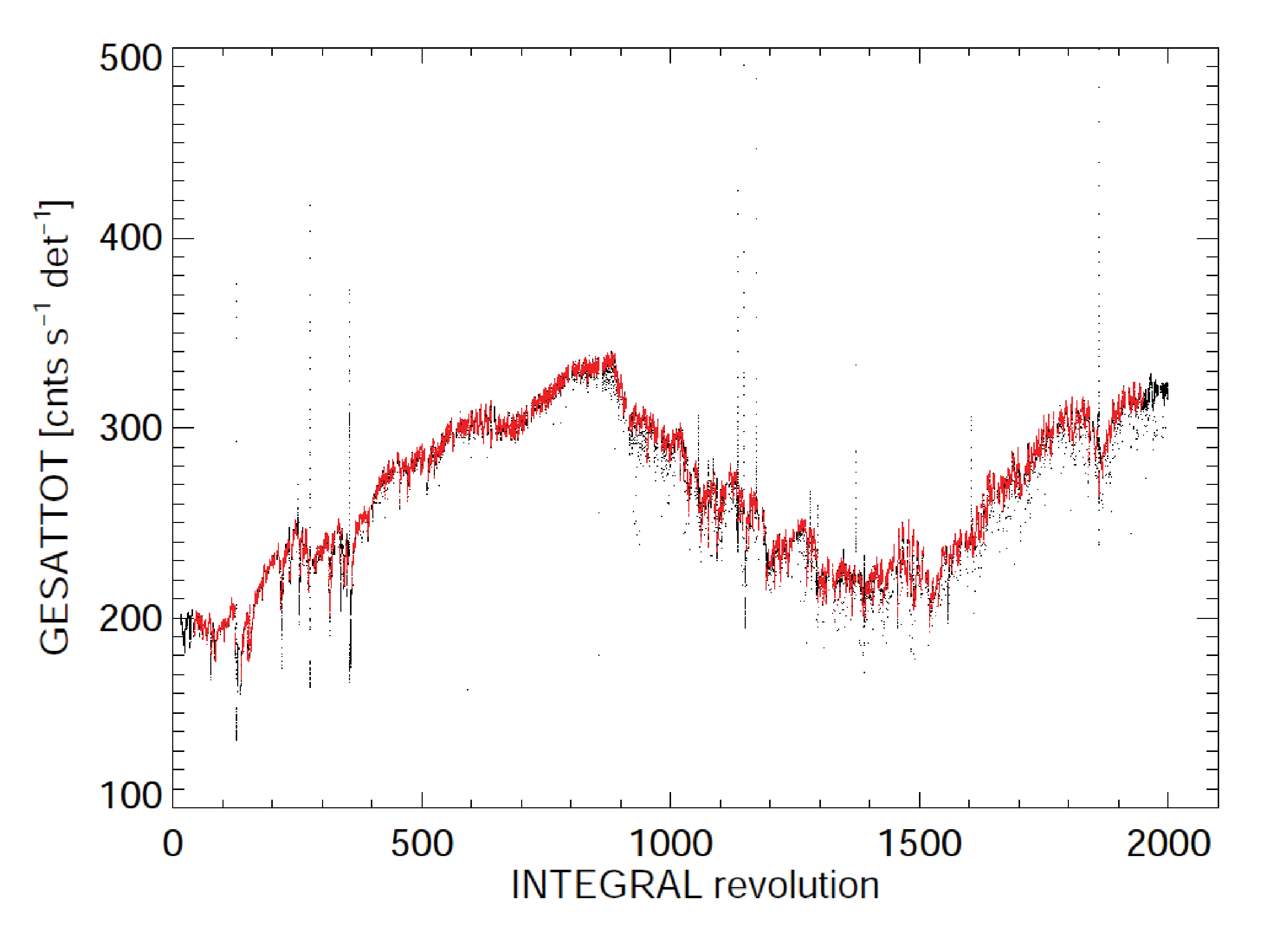}
	\caption{SPI background tracer variations with time from saturated events in the Ge camera (GEDSAT). In the panel, the full SPI data base is shown in black, and the chosen data based on our selection criteria in red.}
\label{fig:gedsattracer}
\end{figure}

\begin{figure}
	\centering
	{\includegraphics[width=5cm]{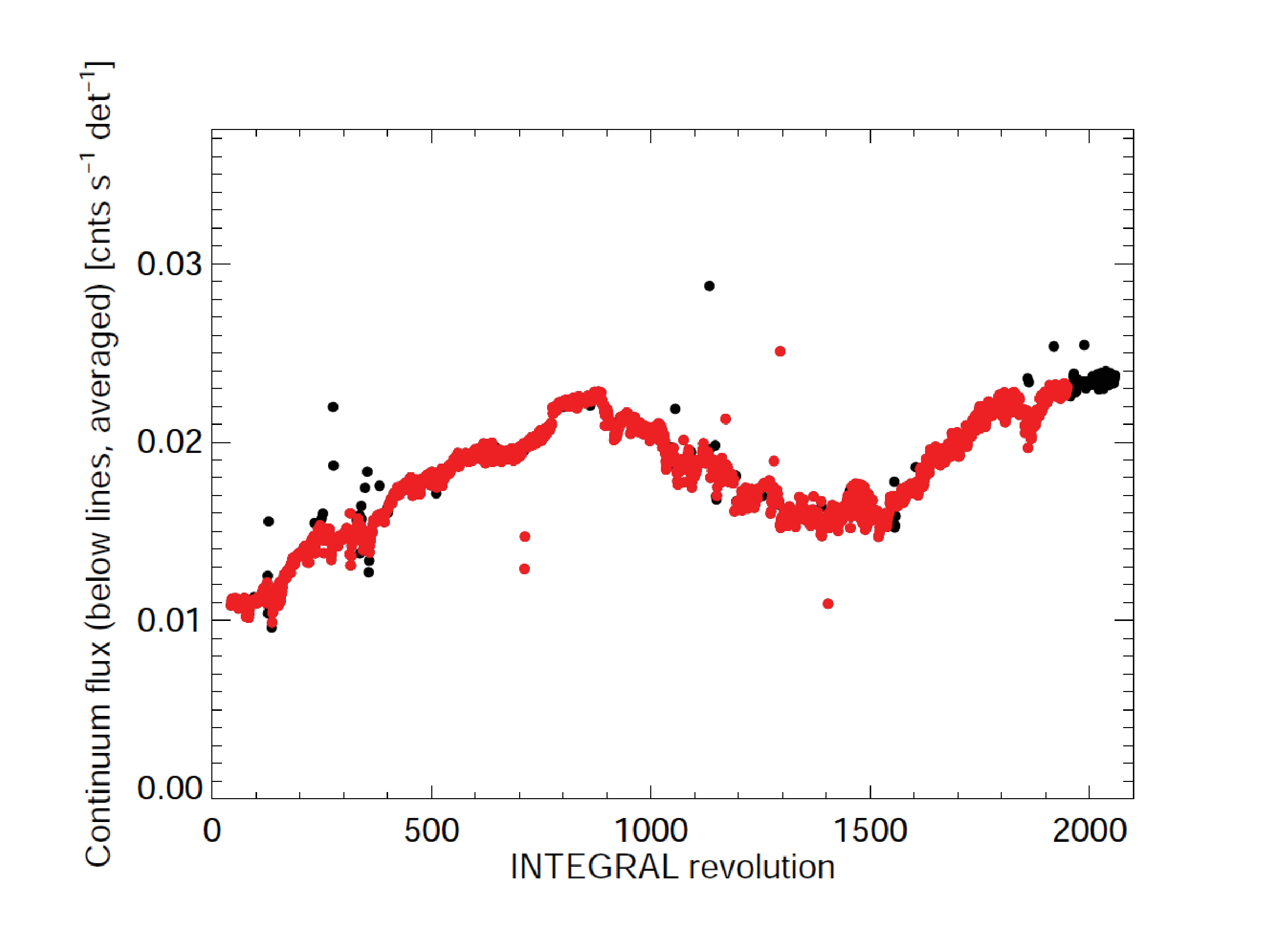}}
    {\includegraphics[width=5cm]{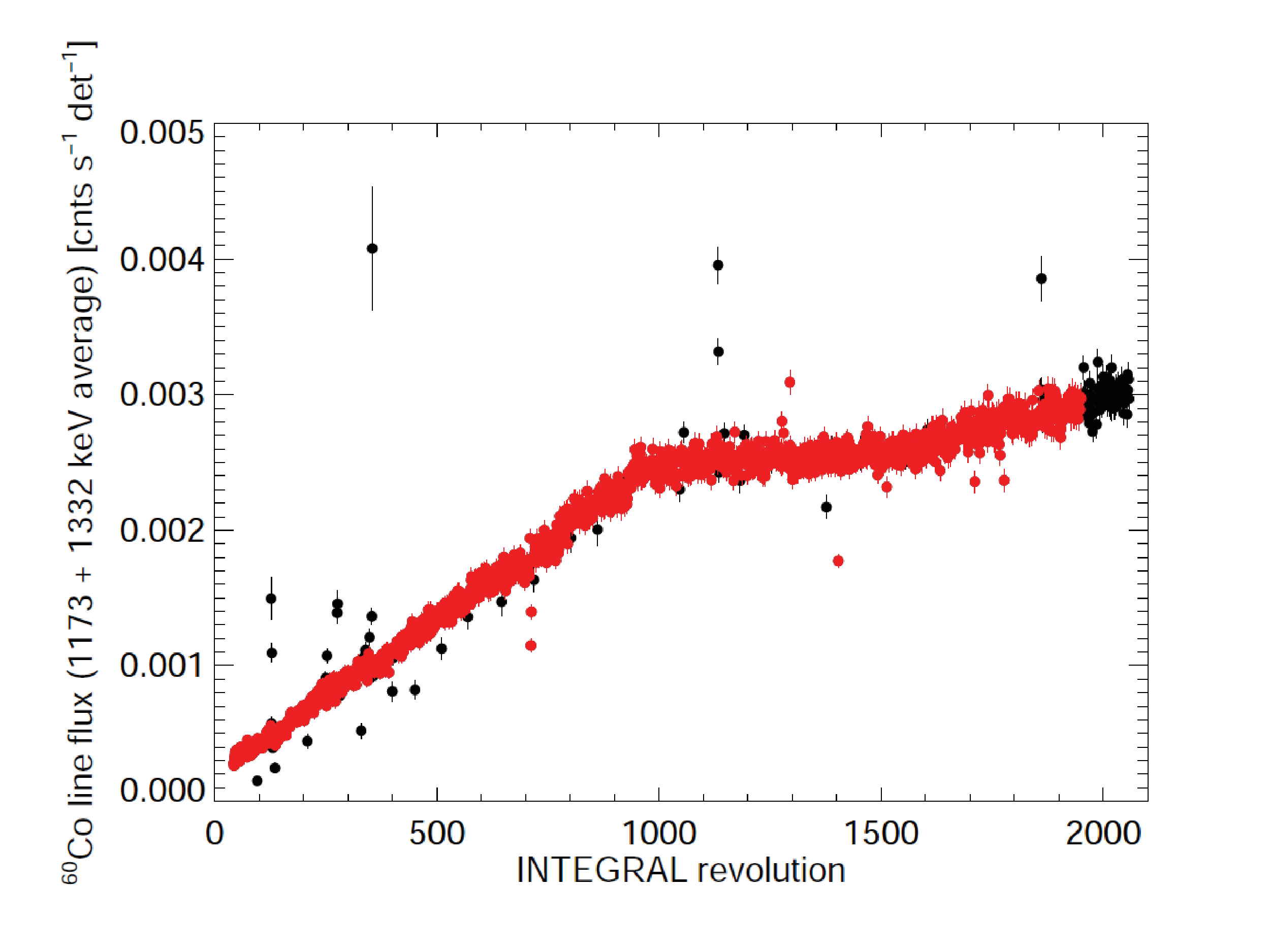}}
	{\includegraphics[width=5cm]{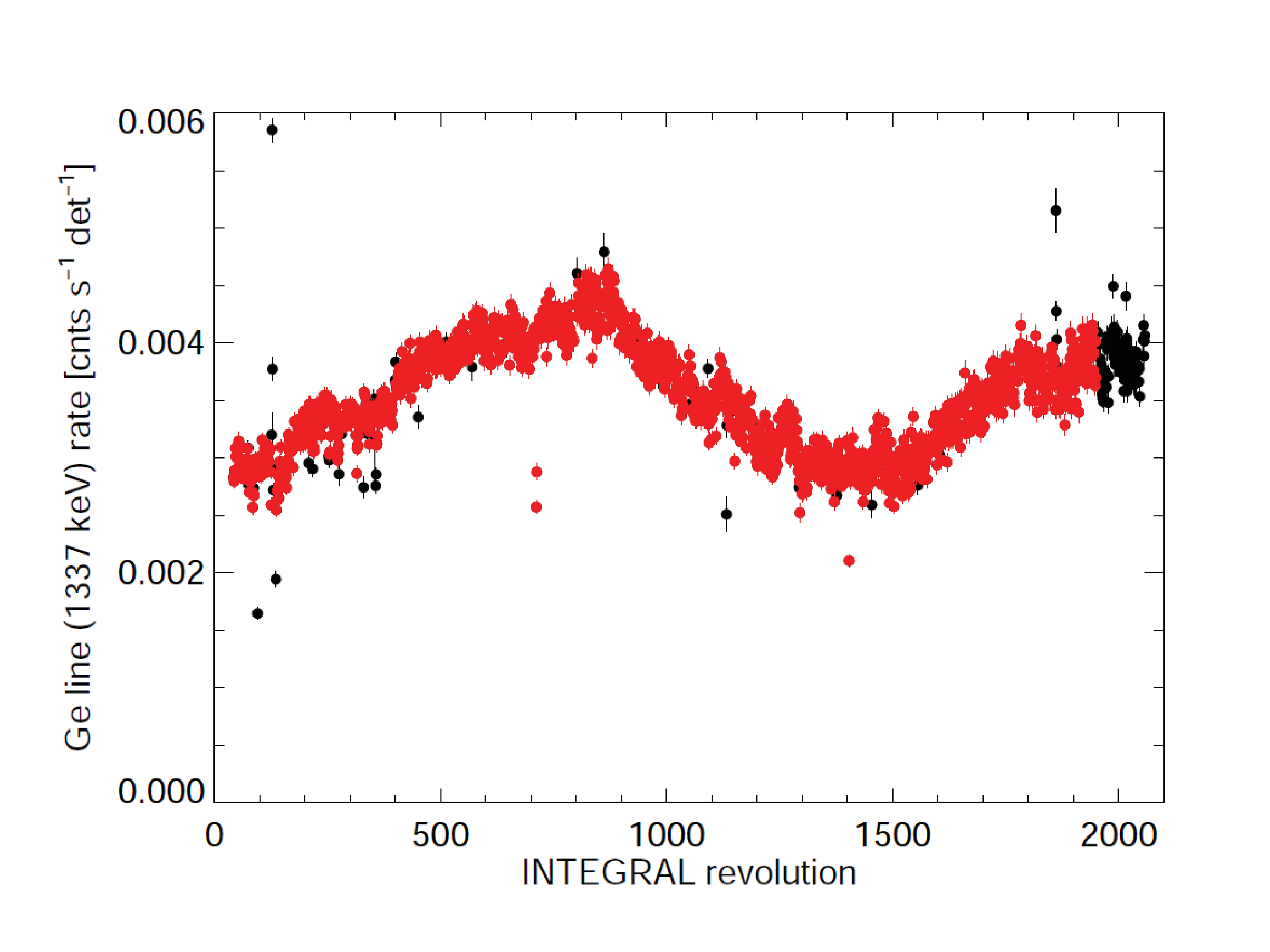}}
	\caption{SPI background continuum and line components relevant in the spectral regime of \Fe lines: (left) the continuum count rate, (middle) $^{60}$Co activation, (right) and a Ge activation line. In each panel, the full SPI data base is shown in black, and the chosen data based on our selection criteria in red.}
\label{fig:bgmodel}   
\end{figure}

\subsection{\Fe Background characteristics}\label{sec:bg_char}

Much of the radiation from instrumental background is promptly emitted directly from CR impacts. Background components arise from radioactive isotopes produced
by the CR impacts. Local radioactivity in the spacecraft and instruments themselves
thus will generate both broad continuum background emission and
narrow gamma-ray lines from long and short-lived radio-isotopes. Varying
with energy, background components may exhibit complex time
variability due to their origins from more than one physical
source (Weidenspointner et al. 2003). Background modelling by using the entire mission spectroscopy history has been established recently (Siegert et al. 2019).

For the energy bands studied here, we follow this general approach, and model the background by fitting two components, one for the continuum and one for instrumental lines.
From the mission spectral database (Diehl et al. 2018), we construct these background models at fine spectral precision. Then, we allow for adjustment of its overall normalisation in intensity, which accounts for the fact that the (small) celestial signal that had been part of the mission data used for this database now needs to be separated out. The instrument records several detector triggers which have sufficiently-high count rates, such as the one of onboard radiation monitors, the SPI anti-coincidence shield count rate, and the rate of saturating Ge detector events (events depositing $>$8 MeV in a Ge detector, GeDSat). These count rates reflect the current cosmic-ray flux which causes the prompt instrumental $\gamma$-ray background, while the long-term trends of radioactive build-up and decays is inherent to the spectral background data base. In this analysis, we use the GeDSat rates as a short-term background tracer and first-order description of the background variations with time scales of pointing-to-pointing, from 800 keV - 2000 keV. This tracing is sufficient in energy bins at or below the instrumental resolution, however in broader energy bins such as used in this work the superposition of the effects from different background lines blended together in a broader bin require re-scalings after suitable time-intervals.

\begin{figure}[!ht]
	\centering
	\includegraphics[width=12cm]{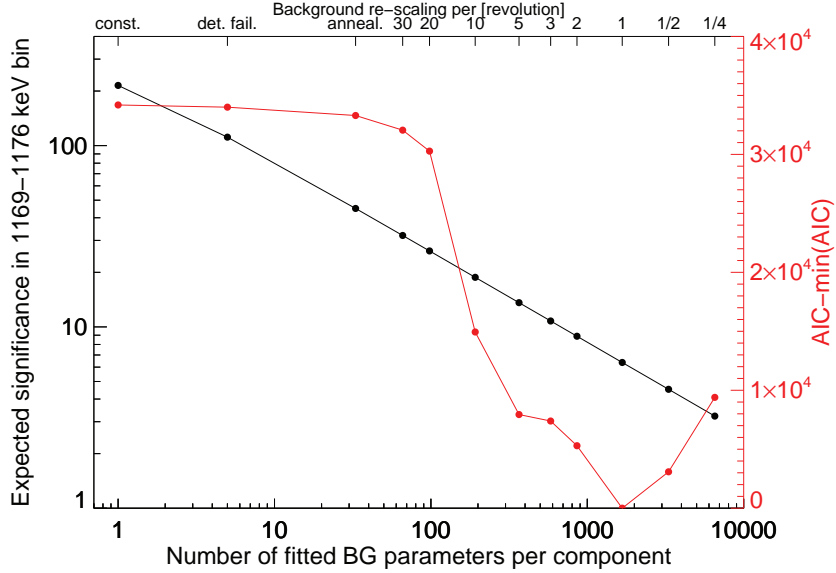}
	\caption{Expected significance (in $\sigma$ units, black points) above a background only description of the data in the 1169 to 1176 keV band, estimated from likelihood tests, BG vs. BG plus source, using different numbers of background parameters (bottom axis), i.e. varying the background on different time scales (top axis: re-scaling per number of INTEGRAL revolutions; or whenever an annealing was performed, a detector failed; or assuming a constant background). The red points show the Akaika Information Criterion ($\mathrm{AIC} = 2n_{par} - 2\ln(L)$, where $n_{par}$ is the total number of fitted parameters, right axis) which was used to find the required number of background parameters to describe the data in this bin sufficiently well. The optimum is found at one INTEGRAL orbit or 1,674 background fit parameters, respectively. The black points show the expected significance in the band with a total flux of $4 \times 10^{-4}\,\mrm{ph\,cm^{-2}\,s^{-1}}$ as a function of fitted background parameters. See text for details.}
	\label{fig:aicanalysisbgscalingfe1}
\end{figure}

The coefficients $\beta_{b,t}$ for background normalisations are therefore allowed time dependent, to cater for such effects, and for different
background normalisations for each camera configuration of
19/18/17/16/15 functional detector elements, as well as for possible variations on time scales shorter than our background model was built.

In Fig.\,\ref{fig:aicanalysisbgscalingfe1}, we show the optimal re-normalisation time scale for the \Fe energy bin including the 1173\,keV line. We find an optimum fit when re-normalising background at a time scale of one orbit, which corresponds to 1,674 background parameters per component.
As a consistency check, we performed an estimate of the \Fe signal significance of the 1169--1176\,keV band (continuum plus line) as could be expected from earlier measurements (Wang et al. 2007). In this estimation we consider only the inner Galaxy, and assume the COMPTEL \Al map as a tracer for the morphology of \Fe emission, and we adopt a total galaxy-wide flux of $4 \times 10^{-4}\,\mrm{ph\,cm^{-2}\,s^{-1}}$. We then make use of the approach by Vianello (2018), who extended the Bayesian significance estimates of Li \& Ma (1983), and assume herein that our background model parameters are normally distributed. This obtains the black line shown in Fig.\,\ref{fig:aicanalysisbgscalingfe1}, estimating a significance of $6\sigma$ for our optimal background model re-scaling. In case about half of the flux of $4 \times 10^{-4}\,\mrm{ph\,cm^{-2}\,s^{-1}}$ would be degenerate and absorbed in the diffuse $\gamma$-ray continuum component, the line significance would be around $4\sigma$ in our estimate for a single line.
	
The background rescaling investigations in the remaining energy bins show similar optimum time scales, except for the lowest energy bin (800-- 1169 keV), which requires four parameters per orbit.
The number of degrees of freedom is thus 1,591,831 for 1,595,180 data points in the \Fe, \Al, and higher-energy continuum bins, taking into account all fitted parameters for the sky, detector failures and background variations.

Analysing the resulting background variations in the \Fe line bands, we now investigate the candidate origins, based on our detailed spectral analysis of instrumental background with high spectral resolution. The most important background lines are the ones expected from the decay of \nuc{Co}{60} in the instrument and the satellite, at the same line energies because this is the same cascade de-excitation in both cases. This $^{60}$Co background builds up in intensity with time, due to its 5.3\,yr radioactive lifetime, and thus will increasingly contribute to the total measured \nuc{Co}{60} $\gamma$-ray line signal. In addition, there is a strong background line from activated Ge at 1337\,keV, which blends into the high-energy \Fe line at 1332\,keV. This Ge line, however, shows no radioactive build-up as the decay time is of the order of nano-seconds, and hence the count rate in this line closely follows the general activation of backgrounds. All these lines are superimposed onto an instrumental continuum background which is dominated by bremsstrahlung inside the satellite, and also includes Compton-scattered photons and a composite of weaker lines that escape identification in our deep spectral background analysis (Diehl et al. 2018).
In Figs.\,\ref{fig:gedsattracer} to \ref{fig:bgmodel}, we show characteristic background components as they vary with time, as determined from our data set.

Here we focus on the total galactic emission, so that extrapolated estimates from the inner Galaxy towards the full Galaxy only may serve as a guidance, rather than a precise prediction.
Nevertheless, we see that the steadily rising \nuc{Co}{60} background line flux leads to a very shallow increase in the significance of celestial \Fe over time, shallower than \Fe count accumulation alone would suggest. While our previous result with only three years of data (Wang et al. 2007) showed a $4.9\sigma$ signal, in this case using both \Fe lines together and both SE and ME (single and multiple-trigger event types), which is consistent with our expectations, the increase of data by more than 200\,\% in our current dataset would result in only $6\sigma$ significance for SE and the two lines combined. We now also understand the additional background time variation: because the background line rate increases almost linearly (Fig.\,\ref{fig:bgmodel}),  fitting the background requires more parameters than typical for SPI background that follows the solar cycle directly.

\section{Modelling \Fe in the Milky Way}\label{sec:skymodels}

The \Fe signal is too weak to derive the spatial distribution or perform an imaging analysis.
Therefore, we attempt to constrain the size of the \Fe emission region through fitting a parameterized geometrical model, an exponential disk profile, and we determine the scale radius and scale heights. Doing this for all energy bins between 800 and 2000\,keV, we obtain information on how this approach deals with known spatial distributions of $\gamma$-ray emission, thus helping to judge systematics limitations with the \Fe interpretation.
The exponential-disk models in this analysis have been adopted in the following form:

\begin{equation}
\rho(x,y,z) = A \exp\left(-R/R_0\right)\exp\left(-|z|/z_0\right)\mrm{,~with~}R=\sqrt{x^2 + y^2}
\label{eq:expdisks}
\end{equation}

In Eq.\,\ref{eq:expdisks}, $\rho(x,y,z)$ is the 3D-emissivity that is integrated along the line of sight $(l/b)$ to produce maps of sky brightness, with a pixel size of $1^{\circ} \times 1^{\circ}$. These maps are then folded through the coded-mask response to create expected count ratios for all selected pointings.
The normalisation $A$, equivalent to $\beta_s^j$ in Eq.\,\ref{eq:datafit}, is determined (fitted) for a grid of scale radii $R_0$ and scale heights $z_0$ that we test. Here, we use a grid of 16 $R_0$-values, from 500\,pc to 8000\,pc in steps of 500\,pc, times 32 $z_0$-values between 10 and 460\,pc in steps of 30\,pc and between 500 and 2000\,pc in steps of 100\,pc. These models are independently fitted to all seven energy bins, to obtain a likelihood chart for all morphologies tested.
From this, we can determine both the best-fitting flux as well as the best scale dimensions of the emission, plus their uncertainties. Doing this in all our energy bands, we can directly compare the emission size characteristics of \Al versus \Fe, obtaining systematics information from the continuum bands in between (i.e. possible biases for scale dimensions, influenced by the continuum below the lines).

All previous searches for \Fe (Wang et al. 2007; Harris et al. 2005; Bouchet et al. 2015), generally assumed that the \Fe diffuse emission follows the sky distribution of the \Al line. In this study, we explore the morphology of \Fe by comparing with \Al emission as well as continuum emission. We test different tracers of potential candidate sources for \Fe emission, and compare those tracer maps to that of \Al emission as measured and deconvolved from $\gamma$-ray data. This provides an independent judgement of how similar \Fe and \Al are (\S\,\ref{sec:tracerresults}), compared to tracers that may include some of the expected deviations from a strict correlation with \Al. Similar to previous studies (Wang et al. 2009; Siegert 2017), we fit all-sky survey maps from a broad range of different wavelengths to our SPI data.
From this we obtain a qualitative measure which maps are favoured at our chosen energies, respectively. We test a comprehensive set of maps to trace different emission mechanisms which may be related to our SPI data in the different energy bands. The list of tested maps (Tab.\,\ref{tab:tracermaps}) includes also source tracers which may be weakly or not at all related to the candidate \Al and \Fe sources. We use this list of tracers for all or energy bands, in order to reveal degeneracies and systematics, because differences between maps are hard to quantify through fit likelihoods in absolute terms. We also include a background-only fit for reference.  In Tab.\,\ref{tab:tracermaps} we comment on each map briefly to illustrate its main features.

\begin{table}
\caption{The inventory of candidate source tracers, for which we compare how they can represent emission in the 7 energy bands of SPI measurements between 800 and 2000 keV.}
    \centering
	\resizebox{\textwidth}{!}{%
		\begin{tabular}{ll}
			\hline
			Energy & Tracer Type \& Comments \\
			\hline
			408\,MHz & Synchrotron emission of CR e-; mosaic Jodrell Bank/Effelsberg/Parkes (Haslam et al. 1982; Remazeilles et al. 2015) \\
			21\,cm & HI neutral hydrogen, Effelsberg-Bonn HI Survey (EBHIS) (Kerp et al. 2011; Winkel et al. 2016) \\
			1.25--4.9\,$\mrm{\mu m}$ & DIRBE infrared emission from star light of M, K, G stars, 4 individual maps (Hauser et al. 1998) \\
			12--240\,$\mrm{\mu m}$ & IRAS infrared emission from dust, 6 individual maps (Hauser et al. 1998) \\
			$380$--$672\,\mrm{nm}$ & Optical emission, all sky mosaic from $>$3000 CCD frames (Mellinger 2009) \\
			656\,nm & $\mrm{H\alpha}$ emission, partly ionised interstellar gas, star forming regions (Haffner et al. 2016) \\
			1.809\,keV & \Al decay emission from , massive star groups, COMPTEL (Plueschke 2001), and SPI (Bouchet et al. 2015) \\
			1--30\,MeV & COMPTEL MeV $\gamma$ rays, CR-ISM interactions (Sch\"onfelder et al. 1993; Strong et al. 1994) \\
			$>$100\,MeV & EGRET  $0.1$--$30\,\mrm{GeV}$ band; CR-ISM interactions (Hartman et al. 1999) \\
			1--3\,GeV & Fermi/LAT, CR-ISM interactions; high-energy sources (Atwood et al. 2009) \\
			0.25--1.5\,keV & ROSAT all sky survey, hot ISM, X-ray binaries, 3 individual maps (Snowden et al. 1997; Voges et al. 1999) \\
			14--150\,keV & Swift/BAT hard X-ray sources, X-ray binaries, point-like, 7 individual maps (Krimm et al. 2013) \\
			30--857\,GHz & Planck radio bands, 9 individual maps, synchrotron emission; individual sources (Planck collaboration 2016) \\
			30--857\,GHz:  AME & Anomalous Microwave Emission (Planck collaboration 2016) \\
			30--857\,GHz:  CMB & Cosmic Microwave Background (Planck collaboration 2016) \\
			30--857\,GHz:   CO & $J(1\rightarrow0)$ emission at 150\,GHz (Planck collaboration 2016) \\
			30--857\,GHz:   Dust & - (Planck collaboration 2016) \\
			30--857\,GHz:   FreeFree & Bremsstrahlung emission  (Planck collaboration 2016) \\
			30--857\,GHz:   Synchrotron & -  (Planck collaboration 2016) \\
			30--857\,GHz:   SZ-effect & Sunyaev-Zeldovich effect (Planck collaboration 2016) \\
			30--857\,GHz:   X-lines & Strong non-CO lines in the centre of the Galaxy (Planck collaboration 2016)  \\
			\hline
	\end{tabular}}
	\label{tab:tracermaps}
\end{table}

\begin{figure}
	\centering
	\includegraphics[width=0.7\linewidth]{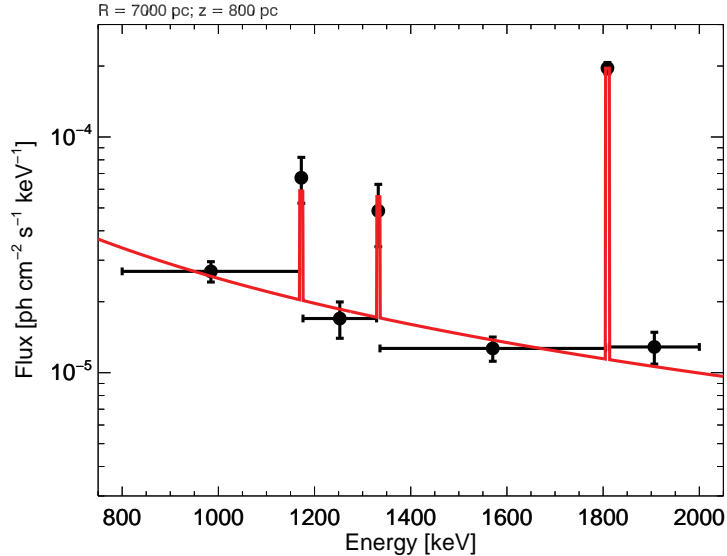}
	\caption{Spectral intensities (black) obtained from the fit to an exponential-disk model with $R_0 = 7$\,kpc and $z_0 = 0.8$\,kpc. The fitted total model, Eq. 4, is shown in red.}
	\label{fig:specalfitnew}
\end{figure}

\section{Results}\label{sec:results}

From independently fitting spatial emission models to SPI data for each of our seven energy bands, we obtain intensity values for the celestial emission detected in each of these bands, for the same adopted spatial distribution model.  In Tab.\,\ref{tab:chi2}, we presented the $\chi^2$ values for seven energy bands in the modelling fittings. So that the present fits are reliable, and the background model is acceptable.

For further discussion and analysis, we fit each extracted set of sky-intensity values with
\begin{eqnarray}
F(E;C_0,\alpha,F_{60},F_{26}) & = & C_0 \times \left(\frac{E}{1000\,\mrm{keV}}\right)^{\alpha} + \\\nonumber
& + & F_{60} \times (T(1172.5\,\mrm{keV},7\,\mrm{keV}) + T(1332.5\,\mrm{keV},7\,\mrm{keV})) + \\\nonumber
& + & F_{26} \times T(1809.0\,\mrm{keV},8\,\mrm{keV})\mrm{,}
\label{eq:specmodel}
\end{eqnarray}
where $C_0$ is the continuum flux density nomalisation at 1000\,keV, $\alpha$ is the power-law index, and $F_{60}$ and $F_{26}$, respectively, are the integrated fluxes as derived from tophat functions, $T(E_0,\Delta E)$, centred at $E_0$ with bin width $\Delta E$.
We link the parameters of the two \Fe lines as they are expected to reflect the same incident flux in intensity and width. The lines are expected to be somewhat broadened above instrumental line widths by 0.1--0.2\,keV due to large-scale galactic rotation of sources  (Wang et al. 2009, Kretschmer et al. 2013), and the instrumental line width would be $\sim2.8$\,keV around the \Fe lines and $\sim3.2$\,keV around the \Al line (Diehl et al. 2018). Within the 7\,keV bins for the \Fe lines and the 8\,keV for the \Al line bands thus, $2.9\sigma$ (99.7\,\%) of the expected line fluxes would be contained.

\begin{table}
\caption{The $\chi^2$ values for the analysed set of 7 energy bins from 800 -2000 keV, together with the number of degrees of freedom and the number of fitted parameters, where $\chi^2_P$ refers to Pearson $\chi^2$, $\chi^2_\Gamma$ is modified $\chi^2$, $\chi^2_\Lambda$ is Cstat $\chi^2$ (see Mighell 1999).}
    \centering
	\resizebox{\textwidth}{!}{%
		\begin{tabular}{llllllll}
		\hline
		Energy bin (keV)  & 800-1169 & 1169-1176 &1176-1329 &1329-1336 &1336-1805 &1805-1813 &1813-2000 \\
		\hline
		$\chi^2_P$ & 1,666,345& 1,578,309& 1,586,945 &1,574,026 &1,601,048 &1,579,418 &1,583,406 \\
        $\chi^2_\Gamma$ &1,666,705& 1,577,964& 1,586,890& 1,574,878 & 1,600,992& 1,579,990 &1,583,382 \\
        $\chi^2_\Lambda$  & 1,666,313 &1,582,827& 1,587,166& 1,579,806& 1,601,146& 1,589,270& 1,583,901 \\
        D.o.f.      &    1,591,831& 1,591,831& 1,591,831& 1,591,831& 1,591,831& 1,591,831& 1,591,831\\
        Parameters  & 3,349 &   3,349 &  3,349 & 3,349  & 3,349  &   3,349   &  3,349  \\
		\hline
	\end{tabular}}
	\label{tab:chi2}
\end{table}

\begin{figure}
	\centering
	{\includegraphics[width=0.49\textwidth,trim=0.5in 0.0in 0.5in 0.0in,clip=true]{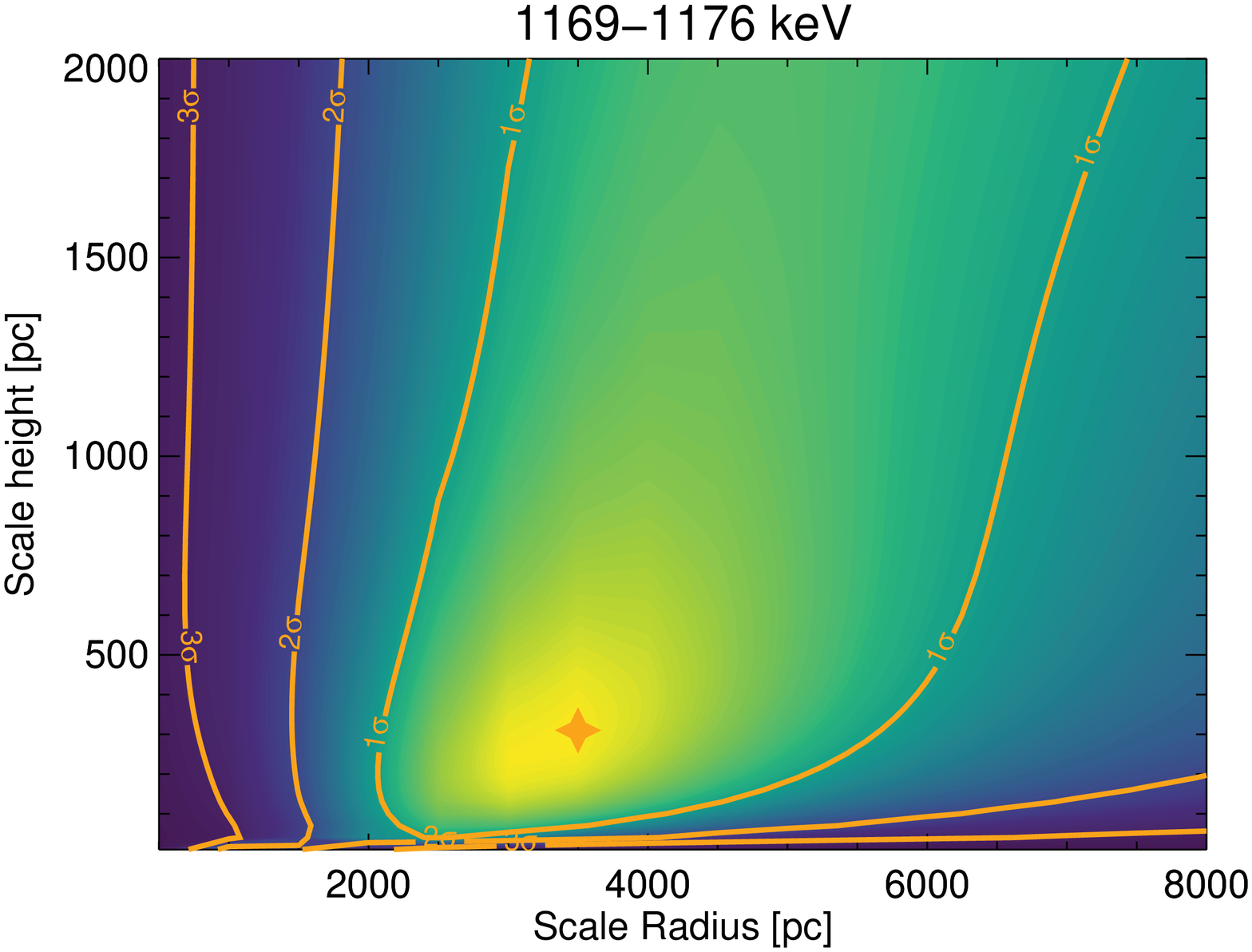}}
	{\includegraphics[width=0.49\textwidth,trim=0.5in 0.0in 0.5in 0.0in,clip=true]{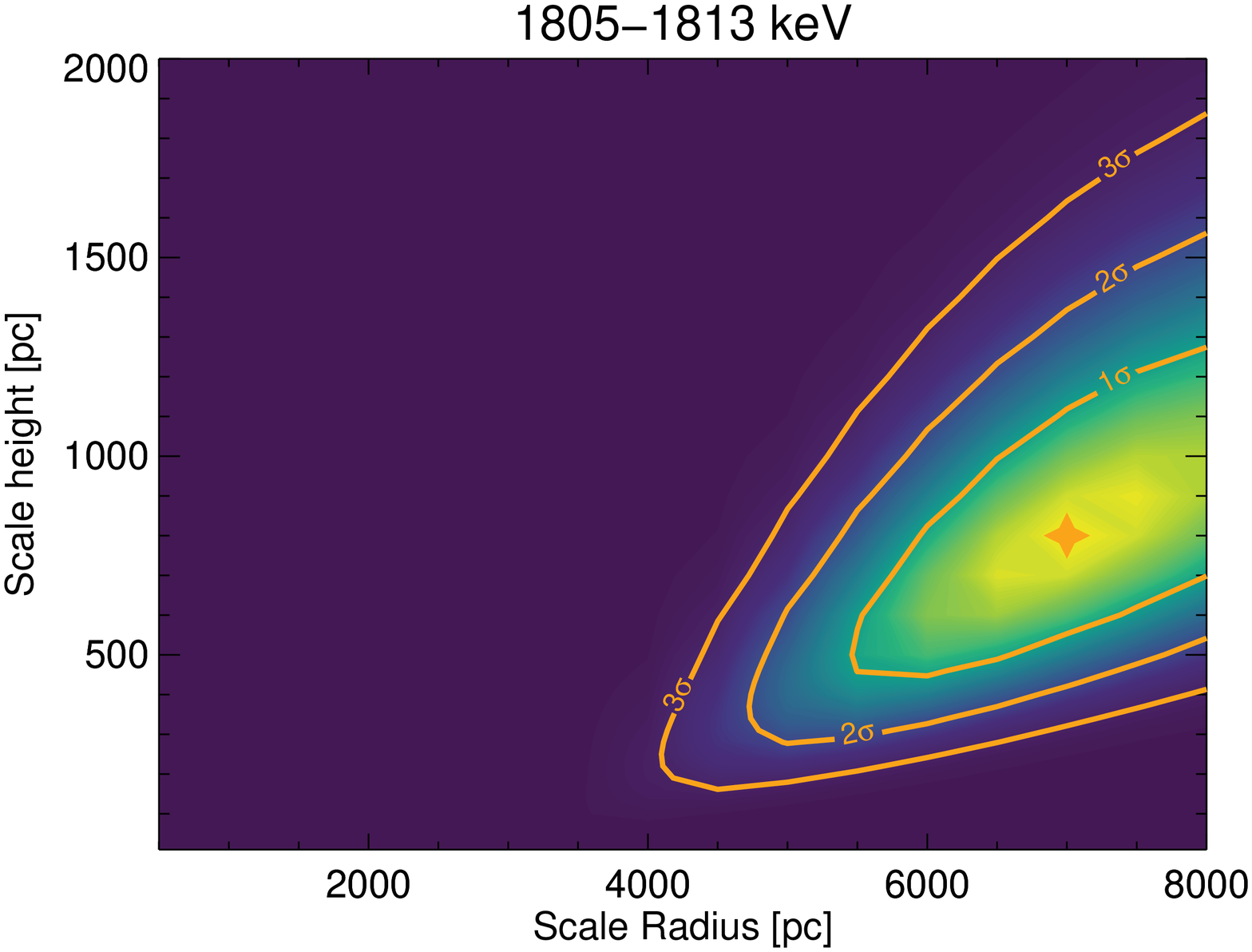}}
	\caption{2D-likelihood profiles for exponential-disk fits to SPI data, as functions of scale radius and scale height. Shown are the 1, 2, and $3\sigma$ uncertainties, and the best-fit value marked with a star symbol. For the \Al line, the best fit is $(R_0/z_0) = (7.0^{+1.5}_{-1.0} / 0.8^{+0.3}_{-0.2})$\,kpc. For the \Fe line, the best fit gives $(R_0/z_0) = (3.5^{+2.0}_{-1.5} / 0.3^{+2.0}_{-0.2})$\,kpc. }
	\label{fig:loglikprofiles}
\end{figure}

\subsection{Characterising the extents of \Fe and \Al emission }\label{sec:expresults}

In Fig.\,\ref{fig:specalfitnew}, we show the 7-band spectral intensities as derived from an exponential disk with scale radius 7\,kpc and scale height 0.8\,kpc, as a typical example.
We selected this as it reflects best-fit dimensions in the \Al line band.
In this example, the continuum is determined as $(2.4\pm0.2) \times (E/\mrm{1000\,keV})^{(-1.3\pm0.2)} \times 10^{-5}\,\mrm{ph\,cm^{-2}\,s^{-1}\,keV^{-1}}$.
The \Fe and \Al line fluxes are $(2.6 \pm 0.6) \times 10^{-4}\,\mrm{ph\,cm^{-2}\,s^{-1}}$ and $(14.4 \pm 0.7) \times 10^{-4}\,\mrm{ph\,cm^{-2}\,s^{-1}}$, respectively, which results in a \Fe/\Al ratio of $(18.3\pm4.4)\,\%$.

Our $16 \times 32 = 512$ scale size grid covers the full range of the Galactic plane, and therefore provides a measure of the emission extents for \Fe and \Al as well as for the continuum emission expected from Bremsstrahlung and inverse-Compton interactions of cosmic-ray electrons.
Each of the  512 exponential disk templates is treated individually in the first place, fitting its parameters without any priors or constraints.
As a result, the absolute fluxes of continuum and lines vary with the emission dimensions.
We find that with larger scale dimensions, the fluxes of continuum and lines increase.
We find no strong variation of the line-to-continuum ratio for either \Fe or \Al.
This supports our assessment that the shape constraints that we derive are consistent and without major bias.

The strong background line at 1337\,keV (see Section 2) could affect the 1332\,keV \Fe line result, which we therefore compare to the more-isolated 1173\,keV line; for our \Fe spatial results, we prefer to rely on the latter line for this reason, while the \Fe/\Al flux ratio uses the data constraints from both Fe line bands combined. In our goal to determine the spatial extent of \Fe emission, we show next to each other for \Fe and \Al in the 1173 and 1809\,keV lines the likelihood contour regions versus scale heights and scale radii  (in Fig.\,\ref{fig:loglikprofiles}). For the \Al emission, we can obtain the characteristic scale radius of $R_0 =7.0^{+1.5}_{-1.0}$ kpc  and scale height of $z_0=0.8^{+0.3}_{-0.2}$ kpc. However, for the \Fe emission lines, the constraints are very poor due to the weak signals, formally resulting in $R_0 = 3.5^{+2.0}_{-1.5}$ kpc, and $z_0= 0.3^{+2.0}_{-0.2}$ kpc. We will use the point source model test to exclude one point source model in the Galactic center (see Section \ref{sec:pointsource}).

\begin{figure}
 	\centering
 	\includegraphics[width=14cm]{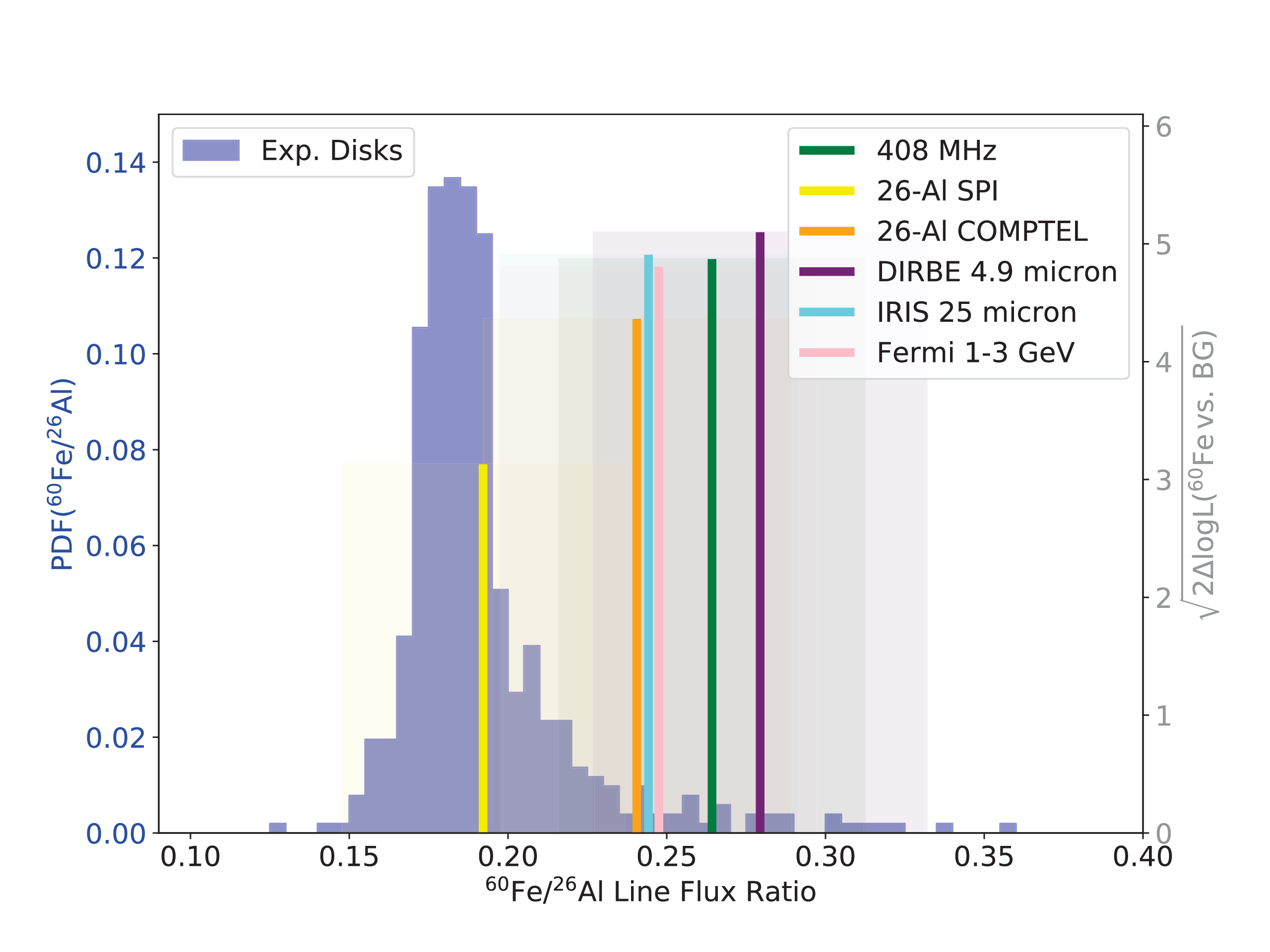}
 	\caption{ \Fe/\Al flux ratio for the grid of exponential disk models (blue, left axis). Including the uncertainties of the fluxes from each spectral fit, the total estimated \Fe/\Al flux ratio from exponential disks is $(18.4 \pm 4.2)\,\%$. Alternative to exponential disks, we also show the flux ratios derived from a set of tracer maps (see section 4.2), as vertical lines according to their significance (right axis), together with their uncertainties as shaded bands. Clearly, these systematically show larger values compared to the exponential disk models. The IRIS (25 $\mrm{\mu m}$) map shows the largest improvement above a background only description for both lines consistently (see Fig. 11), so that a flux ratio estimate from this map serves as a measure of the systematic uncertainty. We find the ratio of $0.24\pm 0.4$ based on the IRAS 25$\mu$m map, suggesting a systematic uncertainty of the order of $6\%$. }
 	\label{fig:fe60toal26rationew}
 \end{figure}

In our goal to exploit maximum information for the \Fe/\Al flux ratio while catering for the uncertainty of the spatial extent of \Fe emission, we include the quality of the fits to SPI data  from this grid of exponential disk model fits in our assessment.
We apply a weighting with the Aikaike Information Criterion (AIC, Akaike 1974), derived from the likelihood and the number of fitted parameters in each energy bin, thus taking the individual measurement and fitting uncertainties of each model map into account.
This yields a \Fe line flux value of $(3.1\pm0.6) \times 10^{-4}\,\mrm{ph\,cm^{-2}\,s^{-1}}$, and a \Al line flux of $(16.8\pm 0.8) \times 10^{-4}\,\mrm{ph\,cm^{-2}\,s^{-1}}$. The \Fe/\Al ratio resulting from this emission-extent averaged analysis is $(18.4\pm4.2)\,\%$. The derived \Al flux for the full Galaxy is also consistent with the previous \Al map study with SPI data (Bouchet et al. 2015). In addition, a recent SPI analysis reported the \Al flux values for both the inner Galaxy and the whole sky (Pleintinger et al. 2019): $\sim 0.29\times 10^{-3} \mrm{ph\,cm^{-2}\,s^{-1}}$ for the inner region, and $\sim (1.7-2.1)\times 10^{-3} \mrm{ph\,cm^{-2}\,s^{-1}}$ for the whole sky. We show the \Fe/\Al ratio distribution from all 512 exponential-disk configurations in Fig.\,\ref{fig:fe60toal26rationew}, also indicating characteristic uncertainties.

\subsection{Point source model test for the Galactic plane}\label{sec:pointsource}

\begin{figure}
	\centering
	{\includegraphics[width=13cm]{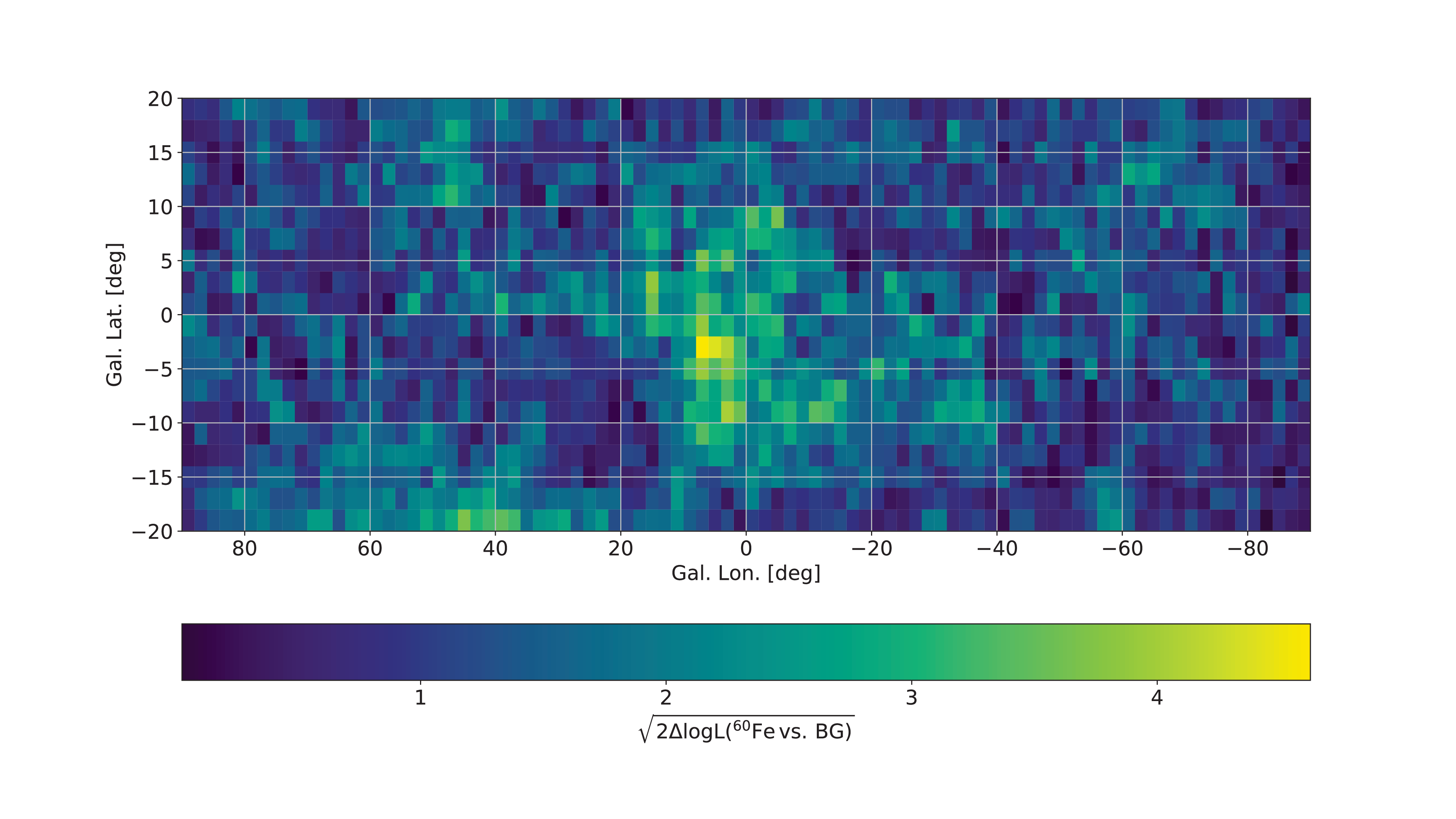}}
	{\includegraphics[width=13cm]{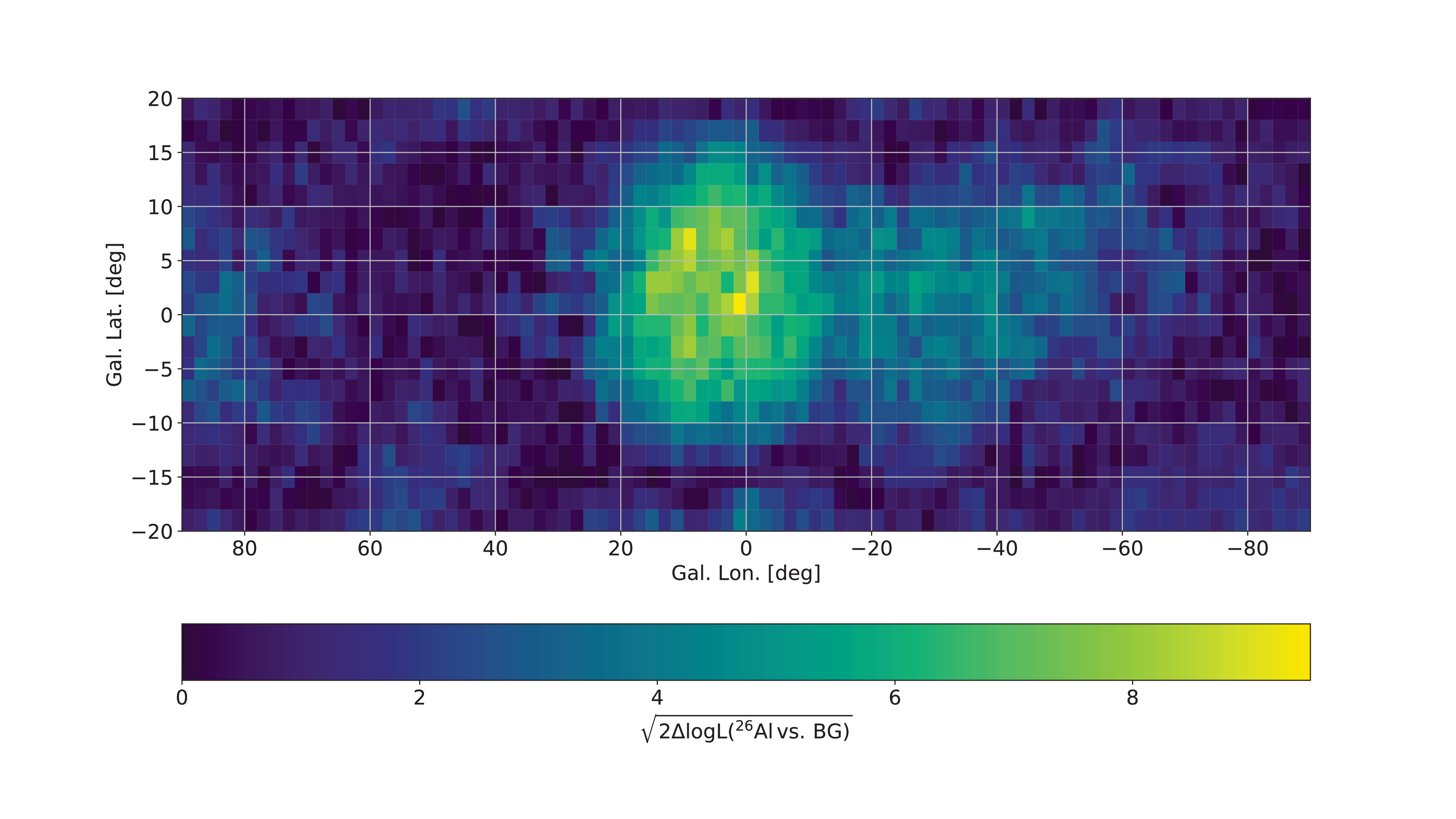}}
	\caption{Summary of scanning the inner part of the Galactic plane for both \Fe (top) and \Al (bottom) line emissions ($-90^\circ <l<90^\circ$, $-20^\circ<b<20^\circ$) with individual point sources, separated by 2 degree each. Each pixel in these visualisations represents one complete likelihood ratio test of BG only vs. BG plus point source, i.e. it includes all fitted parameters of the complete data set with one additional sky component, here modelled as a point source at Galactic coordinates (l/b). Each point is independent from all other points, as they represent another likelihood ratio test, and thus may not be interpreted as ¡®linked to each other¡¯. The particular choice of fitting a single point source at individual positions stems from the fact that, within 3$\sigma$ uncertainties, the exponential disk model extents (see Fig. 7) are indistinguishable from a point source at the Galactic centre. Thus the opposite extreme of having only one or more point source containing all the flux is tested with this procedure. }
    \label{fig:pointsourcescan}
\end{figure}

In this section, we aim to constrain the morphology of the weak \Fe in another way. We produced a catalogue of point source locations with $91\times 21 =1911$ entries between $l=-90^\circ - 90^\circ$ and $b=-20^\circ - 20^\circ$, in 2 degree steps. Then we used this catalogue to test a point source origin for both \Al and \Fe emission lines in the Galactic plane. In this way, in Fig.\,\ref{fig:pointsourcescan}, we can check if and how extended the \Al and \Fe emissions are. These morphology studies of \Al and \Fe emission distributions suggested the $\gamma$-ray emissions are not attributed to one or several point sources in this region. In positive longitudes, the \Al emission extended to $l\sim 35^\circ$ which may be contributed by Aquila, and at $l\sim 80^\circ$, the emission structure is Cygnus. The truncated structure in positive longitudes will be the influence of the exposure map partially (see Figure 1). In the negative longitudes, the \Al emission extended to $l\sim -75^\circ$, probably Carina. However, these maps should not be interpreted as the real sky distribution map, but can imply that the emission is not point-like. The use of this simplified emission model at different galactic coordinates would yield large residuals in the raw SPI data space. Likewise, the \Fe emission morphology is unknown and could be particularly similar to \Al.

From these number of trials, we can estimate the influence of diffuse or point-like emission by taking into account that a background-only fit would result in a test-statistics defined as:
\begin{equation}
TS = 2\left(\log (L_{BG}) - \log(L_{PS}(l/b))\right)\mathrm{,}
\end{equation}
with $L_{BG}$ and $L_{PS}$ being the likelihoods of a background-only fit and a background plus point-source fit, respectively, for finding a point source by chance at a trial position $(l/b)$, would be distributed as $0.5 \chi^2$ with 3 degrees of freedom (2 position, 1 amplitude), we can compare how much the measured $TS$-distributions deviate. A single point source would have one (or more) high points that at high TS at a particular, sharply defined, $TS$-value. We can see that the \Fe case is deviating from the background-only case in more than a few points and consistently for $TS > 12$. This would be a signature of a diffuse signal for the \Fe emission. For comparison, the \Al line case is shown as well. Therefore, we can exclude a single strong point source in the galactic center as well as somewhere else in this region as the origin of the detected \Fe emission.

\begin{figure}
	\centering
	{\includegraphics[width=16cm]{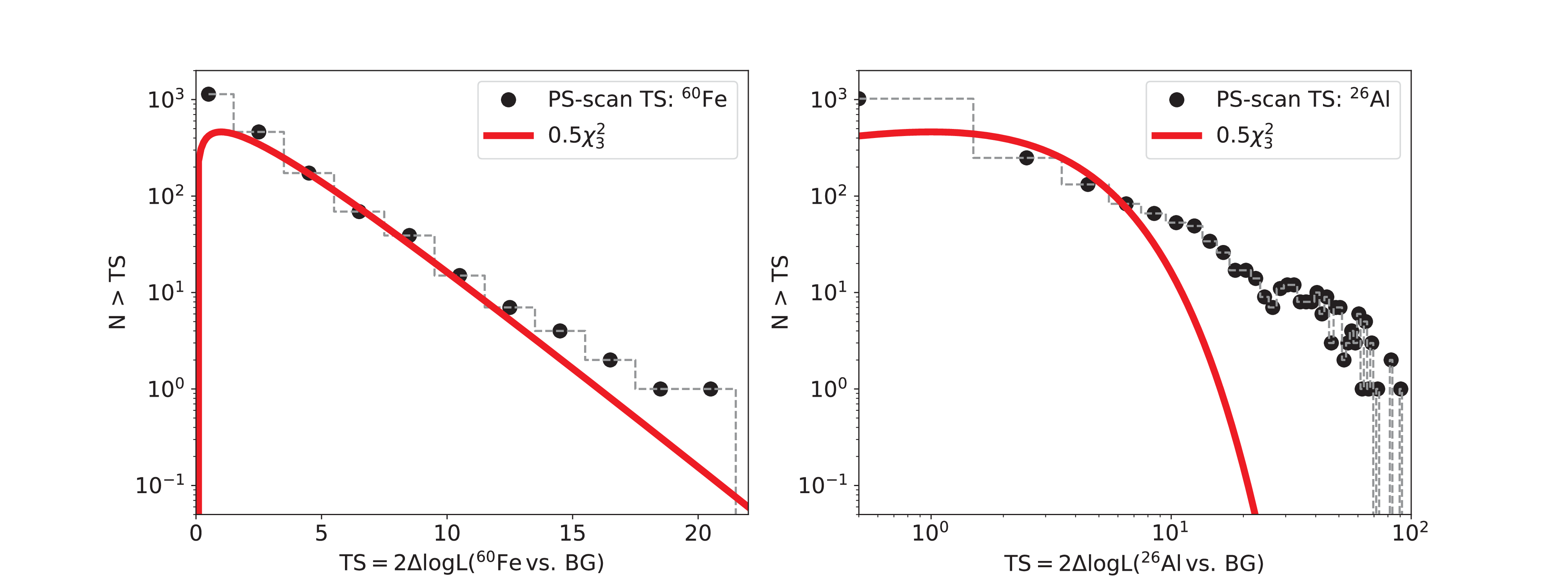}}
	\caption{The test-statistics in the \Fe and \Al lines, together with expectations from BG-only fits, which would be distributed as $0.5\chi^2$ with 3 d.o.f. A single point source would have one (or maybe a few) value at high TS. \Fe is deviating from the background-only case in more than a few points and consistently for $TS > 12$.  }
	\label{fig:teststatitics}
\end{figure}

\subsection{Investigations of different emission tracers}\label{sec:tracerresults}

\begin{figure}
	\centering
	{\includegraphics[width=8cm]{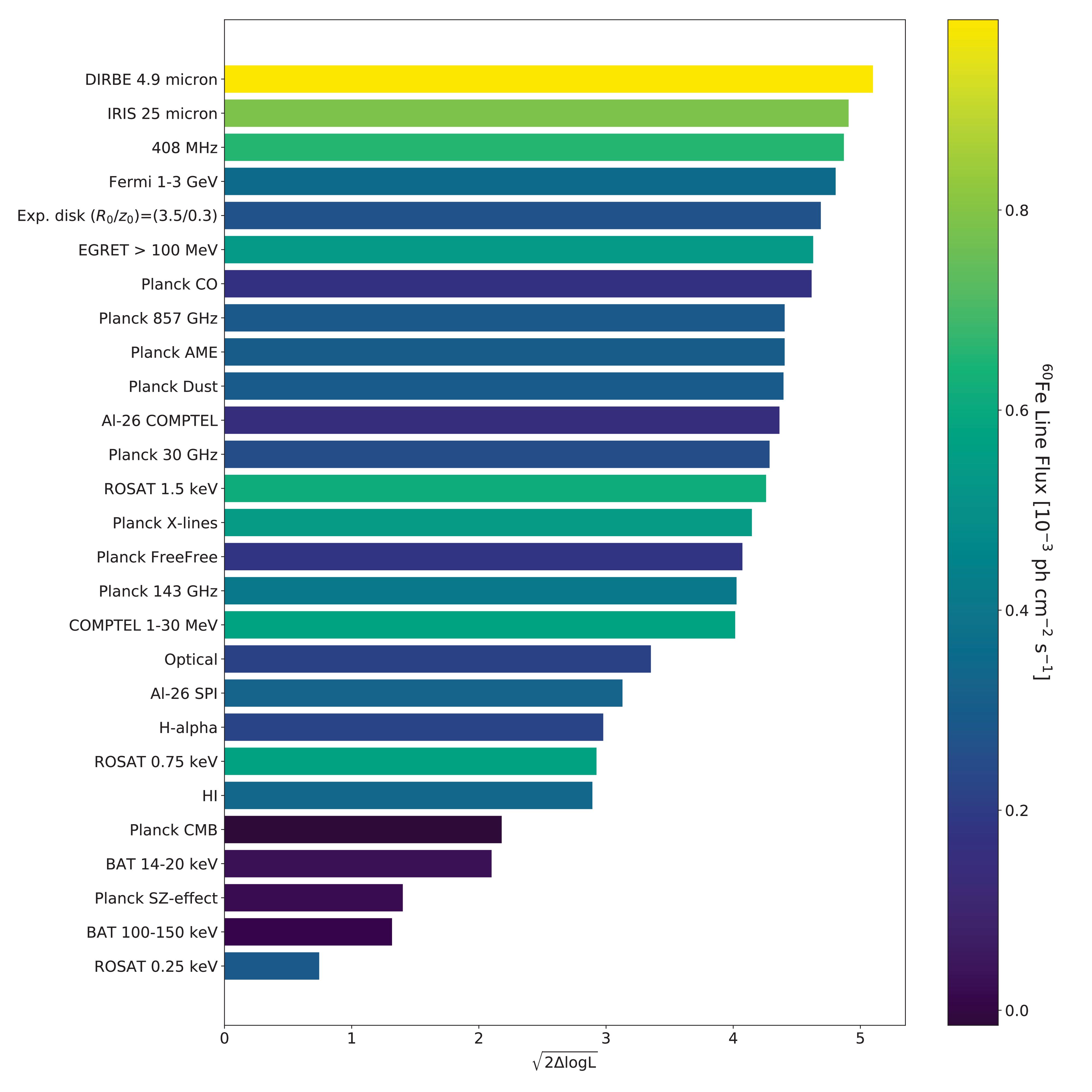}}
	{\includegraphics[width=8cm]{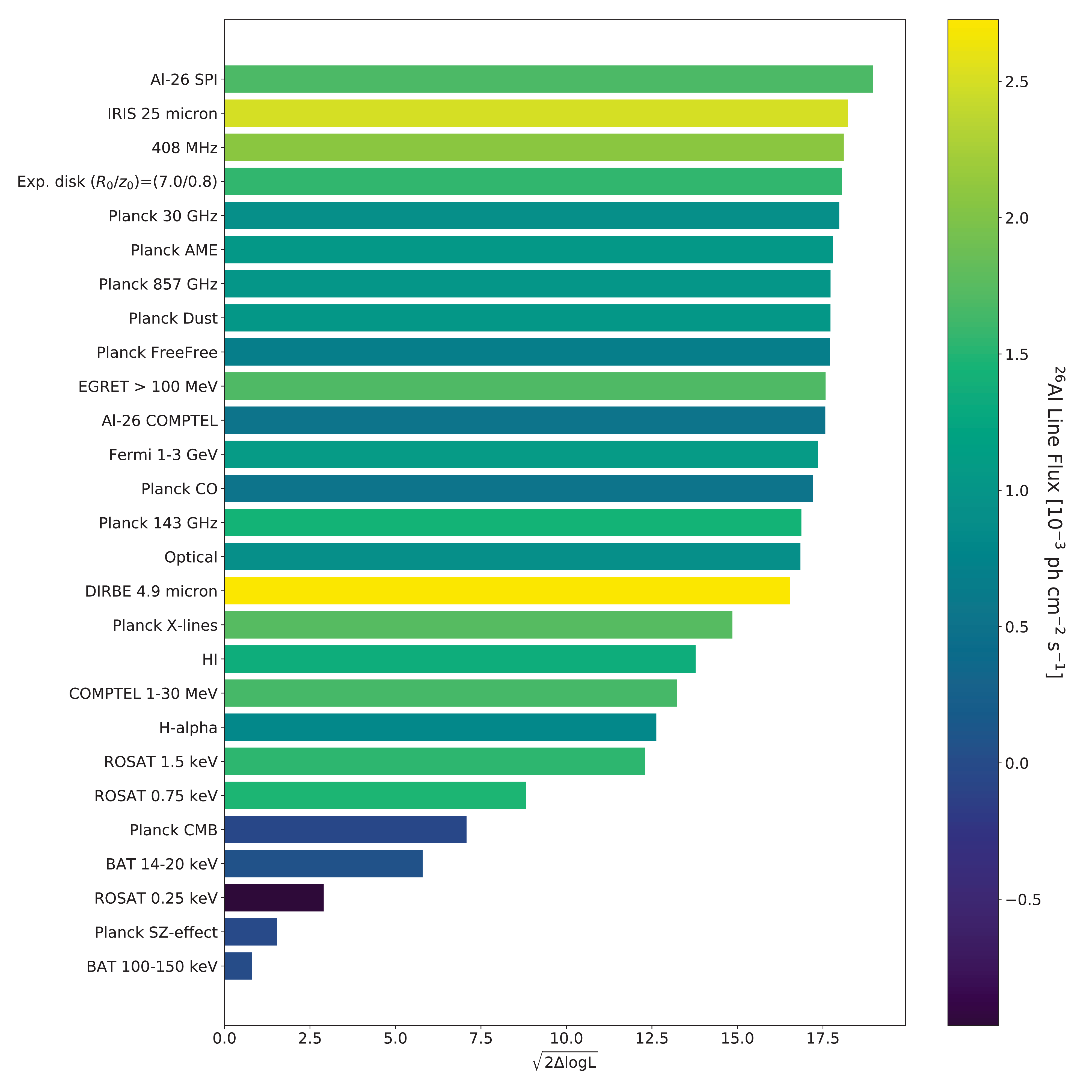}}
	\caption{Likelihood-ratio results for fits of different tracer maps (including the exponential disk maps) of candidate sources to the 1173 keV (left) and 1809\,keV (right) data for \Fe and \Al emissions, respectively. The ratio was derived relative to the background-only fit. The fitted fluxes are given in colour. For the \Fe lines, the best fit sky distribution model is the DIRBE infrared emission map at $4.9\,\mrm{\mu m}$. While for the \Al line, the best fit one is the \Al emission map derived by INTEGRAL/SPI. }
	\label{fig:tracerbars}
\end{figure}

In our effort to investigate the spatial distribution of \Fe emission, we fit the SPI data with a large set of maps representing different source tracers (see Tab.\,\ref{tab:tracermaps}).
These include the 408\,MHz map reflecting cosmic-ray electrons through their radio emission, cosmic-ray illuminated interstellar gas shining in GeV $\gamma$-rays, the COMPTEL and SPI maps reflecting \Al radioactivity, different sets of infrared emission, and X-ray emission maps.
The fit quality of these maps can be compared through their different likelihood ratios, where we normalize to a background-only reference (Fig.\,\ref{fig:tracerbars}).
For the \Al line, as expected, we find again as best-fitting tracer map the SPI \Al line map that had beed derived from a different data set and different analysis method (Bouchet et al. 2015), which supports consistency of our methods. For the \Fe line, the best fitting tracer map turns out to be the DIRBE 4.9$\mu$m map representing emission from small dust grains and star light from mostly M, K, and G-type stars.

From our set of candidate-source tracers, Tab.\,\ref{tab:tracermaps}, we use the likelihood ratio of the combined continuum bins, the two \Fe lines, and the \Al line, each normalized to a background only fit , as a measure for significance of a signal from the sky. We illustrate these signal significance levels and the resulting flux values in Fig.\,\ref{fig:tracerbars}.

For most of the tested maps, such as 408\,MHz, IRAS 25$\mu$m, COMPTEL \Al emission, GeV $\gamma$-ray emission, and CO/dust/free-free emission maps, both the \Al and \Fe radioactive-line bands show a significant detection of a signal from the sky in our SPI dataset, with significance levels of $>16\sigma$ for the \Al line, and $>4\sigma$ for the \Fe lines.
In general, for the cases for which we obtain a significant \Al signal, such as the 408\,MHz map or the COMPTEL \Al map, also \Fe shows a significant signal above the background.
Somewhat surprisingly, however, the SPI \Al map, which fits best at 1809\,keV, is particularly poor in detecting sky emission in the 1173\,keV line band of \Fe.
This may be an indication that \Al and \Fe may indeed have a different emission morphology. In Fig.\,\ref{fig:fe60toal26rationew}, we also presented the \Fe/\Al ratio ranges derived from these tracer maps with the significant detections of both \Al and \Fe emission lines. The \Al all-sky emission maps observed by COMPTEL and SPI gave a ratio range of $0.15-0.24$, and 408\,MHz, IRAS 25$\mu$m and the Fermi $\gamma$-ray emission maps produced a ratio range of $\sim 0.2 -0.3$. The DIRBE $4.9\,\mrm{\mu m}$ emission map can produce a highest detection significance level for \Fe lines, which gave a \Fe/\Al ratio of 0.22 -- 0.32.

For some cases, such as the hard X-ray map from SWIFT/BAT, the soft X-ray (0.25 keV) map from ROSAT, the Planck CMB map, and the SZ-effect map, in both the \Al and \Fe bands we obtain no or at most marginal detections of sky emission. The hard X-ray map (100-150 keV)  is dominated by emission of point sources along the Galactic plane, such as X-ray binaries. Therefore, the non-detection of signals in \Al and \Fe bands would be in line with both the \Al and \Fe emission having a diffuse nature, rather than a strong contribution from such sources.
The Planck SZ effect map follows the distribution of clusters of galaxies, which are mainly located at high Galactic latitudes.
The ROSAT soft X-ray (0.25 keV) map is mostly bright also at high Galactic latitude regions due to the strong soft X-ray absorption by the Galactic plane.
Non-detection of \Al and \Fe emission signals with these two tracer maps therefore is consistent with our belief in origins of \Fe and \Al signals in the plane of the Galaxy and its sources.

\begin{figure}
	\centering
	\includegraphics[width=14cm]{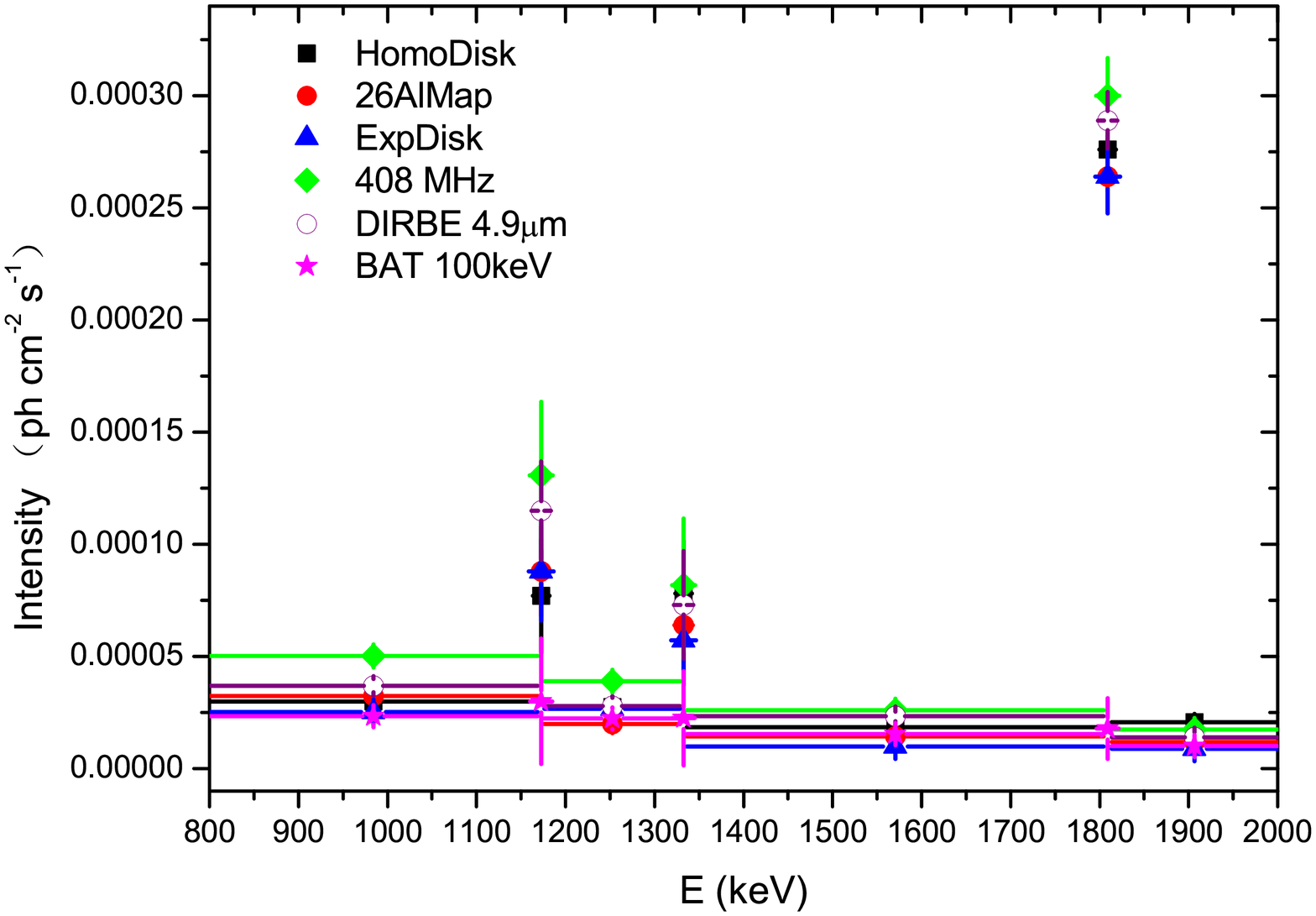}
	\caption{Comparison of the spectra from $800$--$2000\,$keV for different candidate-source tracers in our continuum and line bands. We have included three $\gamma$-ray lines using six typical distribution models: a homogenous disk model (constant brightness along the Galactic plane with scale height 200 pc, see Wang et al. 2009), a COMPTEL maximum entropy \Al emission map (Pl\"uschke et al. 2001), an exponential disk model (scale radius $3.5\,$kpc, scale height $300\,$pc), the 408 MHz map (Remazeilles et al. 2015), the DIRBE infrared emission map at 2.5$\mu$m (Hauser et al. 1998), and a hard X-ray sky map derived by SWIFT/BAT surveys from $100\,$keV--$150\,$keV (bright point sources, Krimm et al. 2013).}
	\label{fig:spectrumphotontracerssample}
\end{figure}

To compare the acceptable fits for the lines and continuum from the set of tracer maps, we show the sample spectra in our energy bands in Fig.\,\ref{fig:spectrumphotontracerssample} from six typical all-sky distribution models, including the diffuse emission maps from observations (408 MHz, \Al $\gamma$-ray emission, infrared emission) and analytical formula (disk models), and point sources based on hard X-ray surveys. This provides an additional check against, or insight towards, possible systematics in our spectral fit results.

\subsection{The diffuse $\gamma$-ray continuum}

The hard X-ray to soft $\gamma$-ray Galactic diffuse emissions include the continuum and $\gamma$-ray lines such as positron annihilation line, \Fe emission lines, and \Al line.
The diffuse continuum emission would originate from several physical processes: inverse-Compton scattering of the interstellar radiation field, bremsstrahlung on the interstellar gas from cosmic-ray electrons and positrons; the neutral pion decays produced in interactions of the cosmic rays with the interstellar gas (see Strong et al. 2010 and references therein), and some unresolved hard X-ray/soft $\gamma$-ray sources.

With the 7-year INTEGRAL/SPI observations, Bouchet et al. (2011) derived the hard X-ray spectrum from 20 keV to 2.4 MeV in the Galactic ridge region,with power-law index of $\Gamma\sim 1.4-1.5$ for the diffuse continuum, and a flux level of $\sim 10^{-5} \rm{ph}\ \rm{cm}^{-2} \rm{s}^{-1} \rm{keV}^{-1}$.
Using the 15-year SPI data covering 800 keV to 2 MeV, we also determine the continuum spectra of the whole Galactic plane from the fittings, which have the average flux level of $\sim (2.0\pm0.4)\times 10^{-5} \rm{ph}\ \rm{cm}^{-2} \rm{s}^{-1} \rm{keV}^{-1}$ with a power-law index of $\Gamma\sim (1.3\pm 0.2)$.
The continuum flux derived in this work is fitted from the whole Galactic plane, rather than only from the inner Galactic ridge (Bouchet et al. 2011). The spectral indices are consistent with each other. We conclude that our broad-band analysis also determines the Galactic continuum emission that underlies the targeted line emissions.

\section{Summary and discussion}\label{sec:discussion}

With more than 15 years of INTEGRAL/SPI observations, we carried out a wide range of spatial model fits to SPI data from 800 -- 2000 keV in seven energy bands, including the line bands for \Fe at 1773 and 1332 keV and for \Al at 1809 keV, as well as wider continuum energy bands around these.
We clearly detected the signals from both \Fe and \Al emissions, as well as diffuse Galactic continuum emission. With only the SE data base, assuming the exponential-disk distribution model, we obtained a detection significance level of \Fe lines of $\sim 5.2\sigma$ with a combined line flux of $(3.1\pm0.6) \times 10^{-4}\,\mrm{ph\,cm^{-2}\,s^{-1}}$, and the \Al flux of $(16.8\pm 0.7) \times 10^{-4}\,\mrm{ph\,cm^{-2}\,s^{-1}}$ for the whole Galaxy above the background continuum. From the consistent analysis approach, with identical data selections, response, and background treatments, we minimise biases or systematics, and obtain a result for the \Fe/\Al flux ratio of $(18.4\pm4.2)\,\%$ based on the exponential disk grid maps. This large-scale galactic value is consistent with a local measurement from deposits of material on Earth (Feige et al. 2018). Since we do not know the real sky distribution of \Fe in the Galaxy, the derived \Fe/\Al flux ratio will depend on the sky distribution tracers. In Fig.\,\ref{fig:tracerbars}, we compared the detection significance levels and flux values for \Al and \Fe using different sky distribution tracers. The best fits for \Al are the the SPI \Al and the IRAS 25$\mu$m maps, while for \Fe, the best one are the DIRBE 4.9$\mu$m and IRAS 25$\mu$m maps. Thus, we can use these best fit maps to constrain uncertainties of the \Fe/\Al flux ratio. If we use the same tracer for both \Al and \Fe, i.e. the IRAS 25$\mu$m map, the \Fe/\Al flux ratio is $0.20-0.28$; then using the different tracers, i.e., the SPI map for \Al and the DIRBE map for \Fe, one finds the ratio range of $0.30 -0.44$.

Using an astrophysically-unbiased geometrical description of a double-exponential disk, we explore a broad range of emission extents both along Galactic longitude and latitude for \Al and \Fe. For \Al emission we find  $R_0 =7.0^{+1.5}_{-1.0}$ kpc and $z_0=0.8^{+0.3}_{-0.2}$ kpc. The \Fe $\gamma$-ray signal is weak and near the sensitivity limit of current $\gamma$-ray telescopes, so that imaging similar to what is obtained for \Al $\gamma$-rays cannot be obtained at present. Formally, the scale radius and height are determined to be $R_0 = 3.5^{+2.0}_{-1.5}$ kpc, and $z_0= 0.3^{+2.0}_{-0.2}$ kpc. We carried out a point source model scan in the Galactic plane ($|l|<90^\circ; |b|<20^\circ$) for both \Al and \Fe emission line cases. The morphology and test-statistics results suggest that the \Fe emission is not consistent with a strong single point source in the Galactic center or somewhere else in the Galactic plane. From our comparison with different sky maps, we provide the evidence for a diffuse nature of \Fe concentrated towards the Galactic plane, which is similar to that of \Al. But it is possible that the \Al and \Fe are distributed differently in the Galaxy.

The ratio of $^{60}$Fe/ $^{26}$Al has been promoted as a useful test of stellar evolution and nucleosynthesis models, because the actual source number and their distances cancel out in such a ratio.  A measurement therefore can help theoretical predictions and shed light on model uncertainties, which are a result of the complex massive star evolution at late phases and related nuclear reaction rate uncertainties. Timmes et al. (1995) published the first detailed theoretical prediction of this ratio of yields in \Fe and \Al, giving a
gamma-ray flux ratio $F(^{60}{\rm Fe})/F(^{26}{\rm Al})= 0.16\pm 0.12$. With different stellar wind models and nuclear cross sections for the nucleosynthesis parts of the models, different flux ratios $F(^{60}{\rm Fe})/F(^{26}{\rm Al})= 0.8\pm 0.4$ were presented (Prantzos 2004). Limongi \& Chieffi (2006) combined their yields for stellar evolution of stars of different mass, using a standard stellar-mass distribution function, to produce an estimate of the overall galactic $^{60}$Fe/$^{26}$Al gamma-ray flux ratio around $0.185\pm 0.0625$.
Woosley \& Heger (2007) suggested that a major source of the large discrepancy was the uncertain nuclear cross sections around the creation and destruction reactions for the unstable isotopes \Al and \Fe which cannot be measured in the laboratory adequately. A new generation model of massive stars with the solar composition and the same standard stellar mass distributes from 13 -- 120 \ms compared yields with and without effects of rotation (Limongi \& Chieffi 2013). For the models including stellar rotation, they determined a flux ratio of $F(^{60}{\rm Fe})/F(^{26}{\rm Al})= 0.8\pm 0.3$. For the non-rotation models, they obtained a flux ratio of $F(^{60}{\rm Fe})/F(^{26}{\rm Al})= 0.2-0.6$; and if one only considers the production of stars from 13 -- 40 \ms, the predicted flux ratio reduces to $\sim 0.11\pm 0.04$. For stars more massive than 40 \ms, the stellar wind and its mass loss effects on stellar structure and evolution contributes major uncertainty in \Fe ejecta production. But these stars may actually not explode as supernovae and rather collapse to black holes, so that their contributions may not be effective and could be ignored in a stellar-mass weighted galactic average. The measured values from gamma-rays suggest that the (generally) higher values from theoretical predictions may over-estimate \Fe and/or under-estimate \Al production. This could be related to the explodability of massive stars for very massive stars beyond 35 or 40~\ms.

\section*{Acknowledgements}
We are grateful to the referee for the fruitful suggestions to improve the manuscript. W. Wang is supported by the National Program on Key Research
and Development Project (Grants No. 2016YFA0400803) and the NSFC (11622326 and U1838103). Thomas Siegert is supported by the German Research Society (DFG-Forschungsstipendium SI 2502/1-1). The INTEGRAL/SPI project has been completed under the responsibility and leadership of CNES; we are grateful to ASI, CEA, CNES, DLR (No. 50OG 1101 and 1601), ESA, INTA, NASA and OSTC for support of this ESA space science mission.

\appendix

\section{Appendix}
\begin{figure}
\centering
\includegraphics[angle=0,width=15cm]{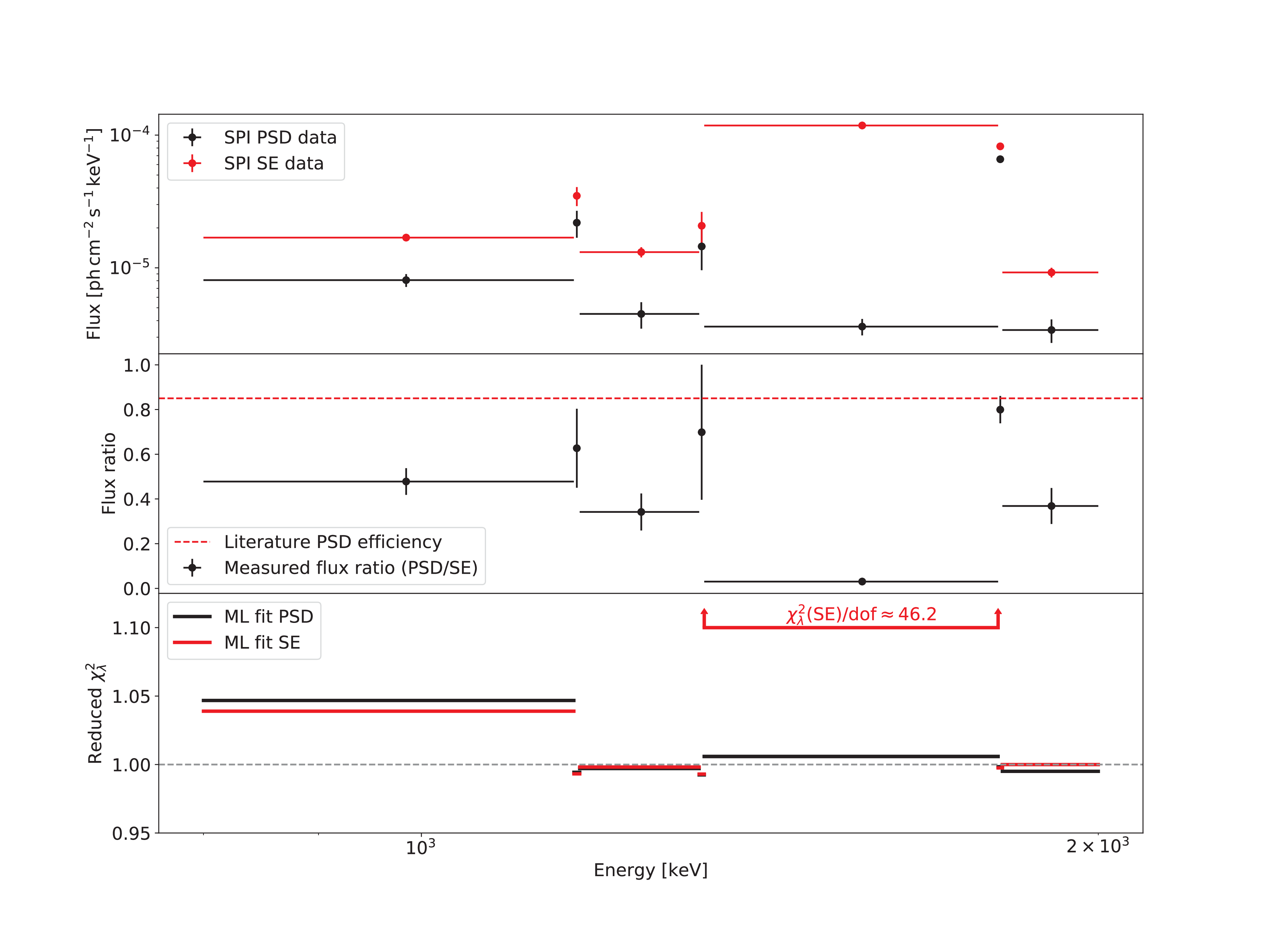}
\caption{A comparison between the fitted broad spectra derived by the SE dataset and PSD dataset from 800 keV - 2000 keV. In the band of 1336 - 1805 keV, the strong electronic noise cannot be suppressed in the case of SE. }
\label{fig:sepsdcomparison}
\end{figure}

In Fig.\,\ref{fig:sepsdcomparison}, we present the spectral examples from 800 keV - 2000 keV for the seven energy bands for both SE and PSD datasets. In the case of SE, the strong electronic noise cannot be suppressed in the band of 1336 - 1805 keV. While, this electronic noise does not affect the spectral counts for the PSD dataset. The reader can also refer to the supplementary information in Siegert et al. (2016), where we have done the test on the pule shape selections. Thus, in this work, we only refer to the PSD dataset.

In the present work, we have tried to constrain the sky distributions of \Fe and \Al lines in the Galaxy, so we determine the gamma-ray spectra of three gamma-ray lines (1173 keV, 1332 keV and 1809 keV) for the entire sky. In the previous work (Wang et al. 2007, 2009), we studied the spectra and fluxes of \Fe and \Al using the maps only covering the inner Galaxy ($-30^\circ<l<30^\circ, -10^\circ<b< 10^\circ$). In the appendix here (see Fig.\,\ref{fig:specinnergal}), we also show the spectral fitting of the broad spectrum with the same map as in Wang et al. (2007, 2009) for a comparison. For the inner Galaxy region, the \Al flux is determined to be $(2.60\pm 0.13)\times 10^{-4} \mrm{ph\,cm^{-2}\,s^{-1}}$, and the combined \Fe flux value is $(4.5\pm 0.8)\times 10^{-5} \mrm{ph\,cm^{-2}\,s^{-1}}$. These flux value levels are consistent with the previous work (Diehl et al. 2006; Wang et al. 2007, 2009; Bouchet et al. 2015). Based on the used COMPTEL \Al map in the inner Galaxy, the \Fe/\Al flux ratio is $0.17\pm0.03$, consistent with the estimates from the full Galaxy, with smaller uncertainties because of the larger average exposure, and increased signal to noise ratio when more flux is actually expected in the analysed region. Of course, for the inner Galaxy, the diffuse gamma-ray continuum has a mean flux of $(4.3\pm 0.6)\times 10^{-6} \rm{ph}\ \rm{cm}^{-2} \rm{s}^{-1} \rm{keV}^{-1}$ at 1 MeV with a spectral index of $1.7\pm 0.3$.

\begin{figure}
\centering
\includegraphics[angle=0,width=15cm]{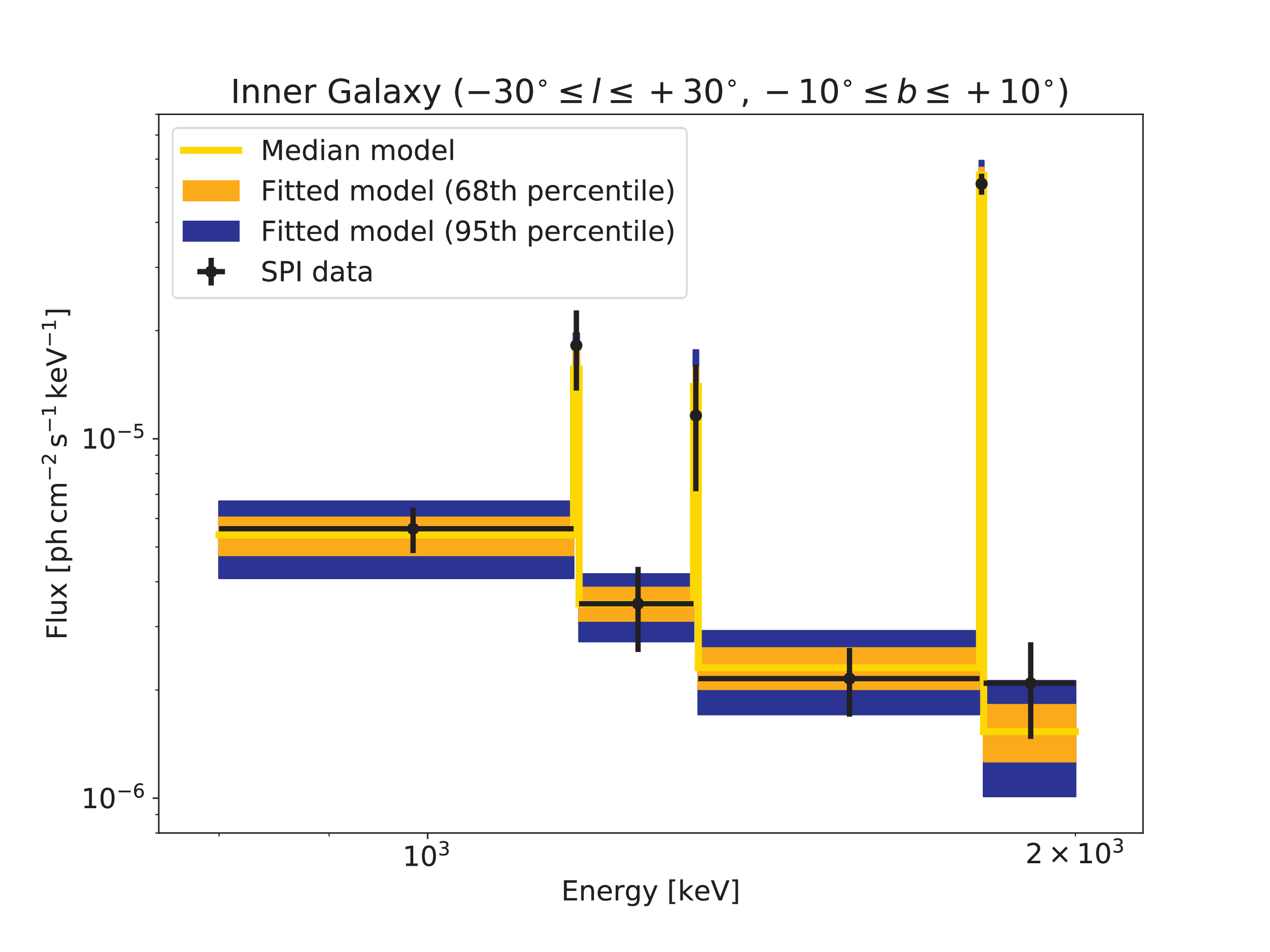}
\caption{The fitted broad spectrum derived from the COMPTEL \Al map only for the inner Galaxy $-30^\circ<l<30^\circ, -10^\circ<b< 10^\circ$. From this fitting, we derive the combined \Fe flux of $(4.5\pm 0.8)\times 10^{-5} \mrm{ph\,cm^{-2}\,s^{-1}}$, and the \Al flux of $(2.60\pm 0.13)\times 10^{-4} \mrm{ph\,cm^{-2}\,s^{-1}}$. These values are consistent with the results in our previous work (Diehl et al. 2006; Wang et al. 2007, 2009).}
\label{fig:specinnergal}
\end{figure}


{}

\end{document}